\documentclass[twocolumn,english,xlinenumbers,aps,twocolumn]{revtex4-1}
\usepackage[T1]{fontenc}
\usepackage[utf8]{inputenc}
\usepackage{color}
\usepackage{babel}
\usepackage{verbatim}
\usepackage{amsmath}
\usepackage{graphicx}
\usepackage[unicode=true,pdfusetitle,
 bookmarks=true,bookmarksnumbered=false,bookmarksopen=false,
 breaklinks=false,pdfborder={0 0 0},pdfborderstyle={},backref=false,colorlinks=true]
 {hyperref}
\hypersetup{
 citecolor=blue}
\usepackage{breakurl}

\makeatletter

\providecommand{\tabularnewline}{\\}

\usepackage{babel}

\graphicspath{{./}{./FIGURES/}} 

\makeatother

\begin{document}

\title{Dynamics of Temporal Localized States in Passively Mode-Locked Semiconductor
Lasers}

\author{C. Schelte$^{1,2}$, J. Javaloyes$^{1}$, S. V. Gurevich$^{2,3}$ }

\affiliation{$^{1}$ Departament de Física, Universitat de les Illes Balears, C/ Valldemossa km 7.5, 07122 Mallorca, Spain}
\affiliation{$^{2}$ Institute for Theoretical Physics, University of Münster, Wilhelm-Klemm-Str. 9, 48149 Münster, Germany}
\affiliation{$^{3}$ Center for Nonlinear Science (CeNoS), University of Münster, Corrensstrasse 2, 48149 Münster, Germany}

%
%
%
%
%
%
%
%
\begin{abstract}
We study the emergence and the stability of temporal localized structures
in the output of a semiconductor laser passively mode-locked by a
saturable absorber in the long cavity regime. For large yet realistic
values of the linewidth enhancement factor, we disclose the existence
of secondary dynamical instabilities where the pulses develop regular
and subsequent irregular temporal oscillations. By a detailed bifurcation
analysis we show that additional solution branches that consist in
multi-pulse (molecules) solutions exist. We demonstrate that the various
solution curves for the single and multi-peak pulses can splice and
intersect each other via transcritical bifurcations, leading to a
complex web of solution. Our analysis is based upon a generic model
of mode-locking that consists in a time-delayed dynamical system,
but also upon a much more numerically efficient, yet approximate,
partial differential equation. We compare the results of the bifurcation
analysis of both models in order to assess up to which point the two
approaches are equivalent. We conclude our analysis by the study of
the influence of group velocity dispersion, that is only possible
in the framework of the partial differential equation model, and we
show that it may have a profound impact on the dynamics of the localized
states.
\end{abstract}
\maketitle

\section{Introduction}

Passive mode-locking (PML) is a well known method for achieving short
optical pulses \cite{haus00rev}. It is achieved by combining two
elements inside of an optical cavity, a laser amplifier providing
gain and a nonlinear loss element, usually a saturable absorber (SA). The
latter favors energetically pulsed emission over continuous wave emission
and, for proper parameters, this combination leads to the emission
of temporal pulses. These impulsions are much shorter than all the
other relevant time-scales, the cavity round-trip $\tau$, the absorber
and the gain recovery times $\tau_{a}$ and $\tau_{g}$, respectively.
Despite having being discovered in 1965 in Ruby lasers \cite{MC-APL-65},
PML is still a subject of intense research, not only due to its important
technological applications \cite{lorenser04,keller96} as high power
sources, especially in vertical cavity surface emitting lasers \cite{haring01,haring02},
see \cite{KT-PR-06} for a review, but also because it involves the
complex self-organization of a large number of laser modes. The PML
dynamics was linked to out-of-equilibrium phase transitions \cite{GP-PRL-02,WRG-PRL-05}
and it can occur without the need of a saturable absorber \cite{S-JSTQE-03,RMW-OE-12}.
The rich PML dynamics can be controlled with time delayed feedback
\cite{JNS-PRE-16} or coherent optical injection \cite{AHP-JOSAB-16}.
In addition, the carrier dynamics in multi-level active materials
as, e.g., quantum dots \cite{RBM-JQE-11,BSR-BOOK-14} leads to even
richer behaviors.

Semiconductors offer unique properties as compared to other materials
and recently, a regime of temporal localization was predicted and
experimentally demonstrated in a semiconductor passively mode-locked
laser \cite{MJB-PRL-14}. It was shown that, if operated in the long
cavity regime, the PML pulses become individually addressable temporal
localized structures (LSs) coexisting with the off solution. This
regime may pave a path towards an optical arbitrary pattern generator
of picoseconds light pulses. Such a functionality would have a large
number of potential applications in different domains, e.g. time-resolved
spectroscopy, pump-probe sensing of material properties, generation
of frequency combs, optical code division multiple access communication
networks \cite{ocdma} and LIDAR \cite{lidar,lidar2}. In this regime,
the temporal interval that corresponds to the cavity round-trip $\tau$
can be seen as a blackboard upon which LSs can be written and erased
at will. Yet, while PML pulses have a duration $\tau_{p}\sim1\,$ps,
they leave in the gain medium a material ``trail'' that follows
their emission. As the gain recovery $\tau_{g}\sim1\,$ns, is slowest
variable, it defines the –effective– duration of the LS, so that the
long cavity regime is only obtained when $\tau\gg\tau_{g}$, which
resulted in a cavity of several meters \cite{MJB-JSTQE-15}. It is
indeed the fast recovery of the gain of the semiconductor that allowed
for the observation of the localization regime. Such a study would
be for instance impractical in fiber or Ti:sapphire lasers \cite{Lederer:99},
for which the gain recovery is several orders of magnitude longer. 

Because of the vast scale separation between the cavity length and
the active gain chip, in our case a vertical-cavity surface emitting
laser (VCSEL) and a resonant saturable absorber mirror (RSAM), the
natural framework for our analysis is that of time-delayed systems
(TDSs) and delay differential equations (DDEs). Interestingly, temporal
LSs were also disclosed in a variety of optical and opto-electronical
time-delayed systems \cite{MGB-PRL-14,MJB-NAP-15,GJT-NC-15,RAF-SR-16}.
Delayed systems have been analyzed from the perspective of their equivalence
with spatially extended systems \cite{GP-PRL-96}, and they have been
shown to exhibit fronts and chimera states \cite{GMZ-PRE-13,LPM-PRL-13,MGB-PRL-14},
see \cite{YG-JPA-17} for a review. It is therefore not entirely surprising
that TDSs may host LSs, which was a result already suggested in \cite{N-PRE-04}.
However, while tempting and intuitive, the ``equivalence'' between
delayed and spatially extended systems sought in the long delay limit
is far from trivial and could so far be formally justified only close
to an Andronov-Hopf bifurcation \cite{GP-PRL-96}. In general, the
non-instantaneous and causal response of the medium implies a lack
of parity in their spatiotemporal representation making the analysis
more involved. While all time-delayed systems are causal and exhibit
some amount of broken parity along the temporal axis, experimental
and theoretical analysis demonstrated that the LSs observed in PML
\cite{MJB-PRL-14} are a most prominent case of parity breaking. These
LSs are particularly stiff multiple timescale objects in which the
optical component and the material ``trail'' differ in extension
by three orders of magnitude, which makes their motion in induced
force fields, induced by, e.g., a modulation of the bias current,
radically different \cite{JCM-PRL-16,CJM-PRA-16} than those found
in parity preserving systems. \cite{MBH-PRE-00,MFH-PRE-02}. To add
to the strong technological relevance in applied photonics of the
temporal localization regime found in PML, the latter was found to
be compatible with spatial confinement, which leads to the theoretical
prediction of a regime of stable three-dimensional light bullets \cite{J-PRL-16}
for realistic semiconductor cavity parameters.

Coarse analytical results regarding the pulse energy only, and preliminary
continuation based upon direct numerical integration allowed finding
some basic estimates of the range of stability for a generic parameter
set. However, a full bifurcation study of the system described in
\cite{MJB-PRL-14} is lacking. A multi-parameter bifurcation study
considering the various design parameters of PML is of high relevance,
as it would inform on the possible mechanisms of instability for these
temporal LSs. The goals of this manuscript are to perform such a bifurcation
analysis and to study the instabilities occurring to the temporal
LSs found in the long delay limit. 

In addition, the multiscale nature of these temporal LSs renders both
their theoretical and numerical analysis difficult. It was shown for
example in \cite{CJM-PRA-16} that an ``equivalent'' master Haus equation
can be used. In this pulse iterative framework, the long tail of the
LS that consists solely in the exponential gain recovery can be truncated,
giving rise to a much more effective numerical approach. While both
models predict very similar waveforms, one can however wonder how
their bifurcation diagrams are consistent one with another. Is is
also our goal to compare the partial differential equation (PDE) model
described in \cite{CJM-PRA-16} with the DDE model of \cite{VT-PRA-05}.
As such, we will compare the bifurcation results obtained in the context
of the time-delayed model, where the LSs were initially discovered,
with those obtained within the framework of an approximately equivalent
spatially extended system, a pulse iterative equation that accounts
for large gain and absorption. 

The paper is organized as follows: In section II, we recall the basic
ingredients of the DDE model \cite{VT-PRA-05}. Section III is devoted
to the bifurcation and the stability analysis of the periodic solutions
found in the long delay limit. For that purpose, we use the continuation
package ddebiftool \cite{DDEBT}. Section IV presents the analysis
of the Haus PDE. In this case, the bifurcation analysis is performed using
the continuation package pde2path \cite{pde2path} and a comparison
is drawn between the two approaches. Finally, our results are summarized
in the conclusion.

\section{Model }

The existence and the dynamical properties of temporal localized structures
in passively mode-locked VCSELs have been theoretically described
\cite{MJB-PRL-14,MJC-JSTQE-16} using the following delay differential
equation (DDE) model \cite{VT-PRA-05} that considers unidirectional
propagation in a ring laser. The equations for the field amplitude
$A$, the gain $G$ and the absorption $Q$ read 
\begin{eqnarray}
\frac{\dot{A}}{\gamma} & = & \sqrt{\kappa}R\left(t-\tau\right)A\left(t-\tau\right)-A,\label{eq:VT1}\\
\dot{G} & = & \Gamma\left(G_{0}-G\right)-e^{-Q}\left(e^{G}-1\right)\left|A\right|^{2},\label{eq:VT2}\\
\dot{Q} & = & Q_{0}-Q-s\left(1-e^{-Q}\right)\left|A\right|^{2},\label{eq:VT3}
\end{eqnarray}
with $R\left(t\right)=\exp\left[\left(1-i\alpha\right)G\left(t\right)/2-\left(1-i\beta\right)Q\left(t\right)/2\right]$,
$G_{0}$ the pumping strength, $\Gamma=\tau_{g}^{-1}$ the gain recovery
rate, $Q_{0}$ the value of the unsaturated losses which determines
the modulation depth of the SA and $s$ the ratio of the saturation
energy of the gain and of the SA sections. We define $\kappa$ as
the intensity transmission of the output mirror, i.e., the fraction
of the power remaining in the cavity after each round-trip. In Eqs.~(\ref{eq:VT1}-\ref{eq:VT3})
time has been normalized to the SA recovery time that we assume to
be $\tau_{sa}=20\,$ps. The linewidth enhancement factor of the gain
and absorber sections are noted $\alpha$ and $\beta$, respectively.
In addition, $\gamma$ is the bandwidth of the spectral filter whose
central optical frequency has been taken as the carrier frequency
for the field. This spectral filter may (coarsely) represent, e.g.,
the resonance of a VCSEL \cite{MJB-JSTQE-15}. In this manuscript,
we will address the bifurcations and the dynamics occurring as a function
of the linewidth enhancement factors $\alpha$ and $\beta$ and of
the gain normalized to threshold $g=G_{0}/G_{th}$, which we define
as our main bifurcation parameters. If not otherwise stated $\kappa=0.8$, $s=30$
and $Q_{0}=0.3$ which corresponds to modulation of the losses of
$\sim26\,\%$. Also, setting $\gamma=10$ and $\Gamma=0.04$, corresponds
to a full width at half maximum (FWHM) of $160\,$GHz for the gain
bandwidth and a carrier recovery time $\tau_{g}=500\,$ps.

\begin{figure}
\centering{}\includegraphics[bb=0bp 20bp 900bp 450bp,clip,width=1\columnwidth]{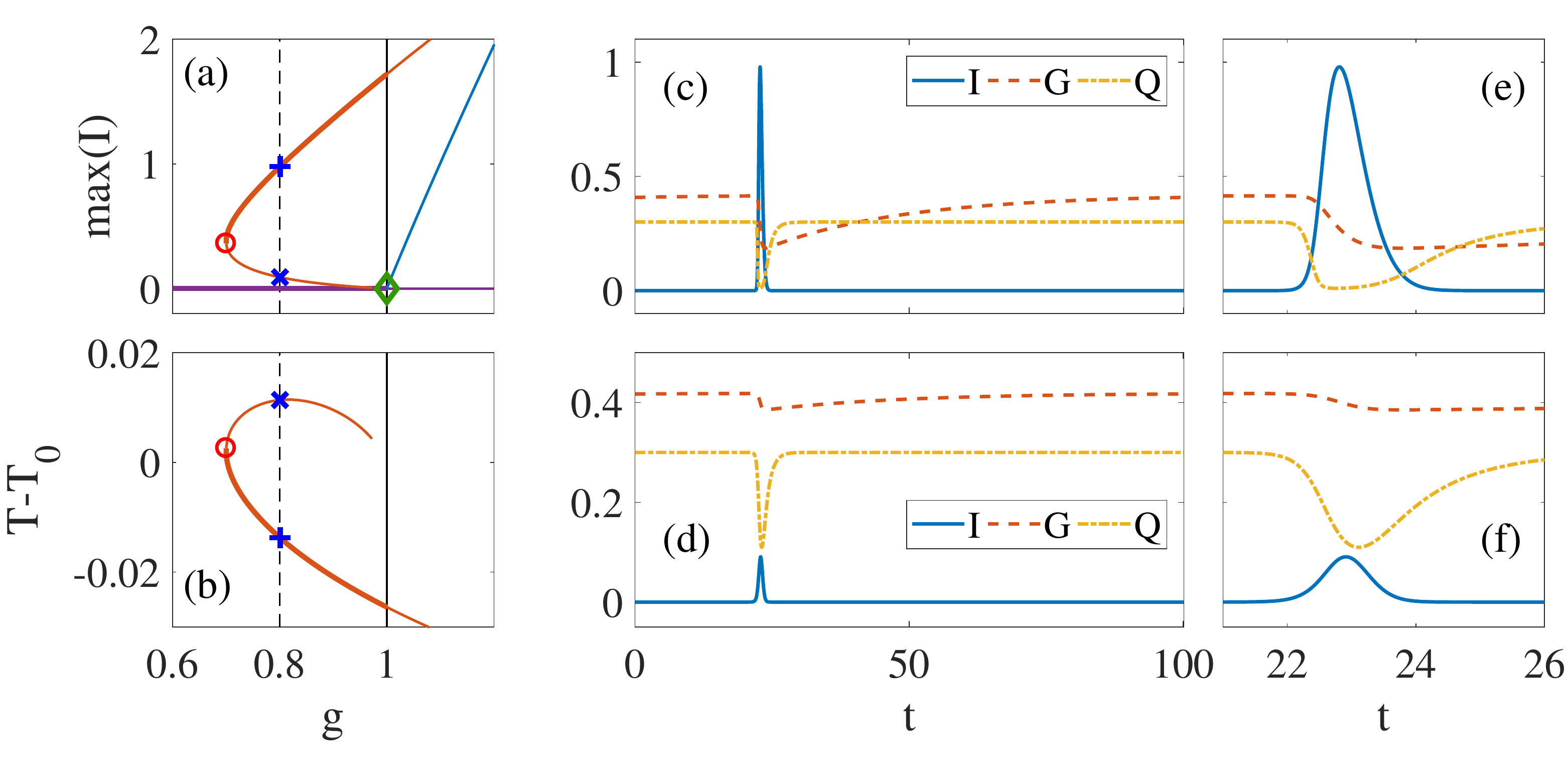}\caption{(Color online) (a,b) Branch of the single temporal LS as a function
of the normalized gain $g$. We represent the maximum intensity (a)
and the deviation of the period of the solution $T-T_{0}$. In (a)
the blue line above threshold is the CW solution with minimal threshold,
whose intensity was multiplied by $10^{2}$ for clarity. Temporal
profiles for the stable solution (c) and the unstable branch (d).
Other parameters are $\left(\alpha,\beta\right)=\left(0,0\right)$.
\label{fig:1}}
\end{figure}

The spatial boundary condition due to the closing of a cavity onto
itself after a propagation length $L$ appears as a time delay $\tau=L/c$
in Eq.~(\ref{eq:VT1}). The latter governs the fundamental repetition
rate of the PML laser. The lasing threshold $G_{th}$ is determined
by the value of $G_{0}$ where the off solution $\left(A,G,Q\right)=\left(0,G_{0},Q_{0}\right)$
becomes linearly unstable. Above threshold, $G_{0}>G_{th}$, multiple
monochromatic solutions $\left(A,G,Q\right)=\left(A_{k}e^{-i\omega_{k}t},G_{k},Q_{k}\right)$
exist \cite{VT-PRA-05}, with an amplitude $A_{k}$ and a frequency
$\omega_{k}$ relative to the filter frequency. If $A_{k}\neq0$,
the modes are defined as the solutions of 
\begin{eqnarray}
\negthickspace\negthickspace\negthickspace\negthickspace\negthickspace1-i\frac{\omega_{k}}{\gamma} & = & \sqrt{\kappa}\exp\left(\frac{\left(1-i\alpha\right)G-\left(1-i\beta\right)Q}{2}+i\omega_{k}\tau\right),\label{eq:SS_CW}
\end{eqnarray}
 complemented with Eqs.~(\ref{eq:VT2}-\ref{eq:VT3}) setting $\dot{G}=\dot{Q}=0$.
Taking the modulus square of Eq.~(\ref{eq:SS_CW}), we find the threshold
condition with $A_{k}\rightarrow0^{+}$,
\begin{eqnarray}
G_{th}^{k} & = & Q_{0}+\ln\left[\frac{1+\left(\frac{\omega_{k}}{\gamma}\right)^{2}}{\kappa}\right]\label{eq:SS_CW1}
\end{eqnarray}
while the modal frequency is given by the ratio of the real and imaginary
parts and reads 
\begin{eqnarray}
\omega_{k}\tau & -\left(\gamma\tau\right)\tan\left[\left(\alpha G_{th}-\beta Q_{0}\right)/2-\omega_{k}\tau\right]= & 0.\label{eq:SS_CW2}
\end{eqnarray}

In the long delay limit, one can safely assume that $\gamma\tau\gg1$
and we can find a good approximation of the frequency of the mode
with the lowest gain threshold $\omega_{0}$. Its expression reads
simply
\begin{eqnarray}
\omega_{0}\tau & = & \Theta
\end{eqnarray}
with $\Theta$ the material induced phase shift per round-trip $\Theta=\left(\alpha G_{0}-\beta Q_{0}\right)/2$.
For this dominant mode, the threshold is $G_{th}^{0}=G_{th}=Q_{0}-\ln\kappa$. 

Temporal LSs appear in TDSs in the long delay limit as periodic orbits
whose period is always slightly larger than the time delay. This deviation
is due to the inertia contained in the structure of a \emph{differential}
equation like Eq.~(\ref{eq:VT1}). In our case, the physical interpretation
of this reaction time is stemming from the finite bandwidth of the
filter. The nominal period of the orbits in a PML laser described
by Eqs.~(\ref{eq:VT1}-\ref{eq:VT3}) is defined as $T_{0}=\tau+\gamma^{-1}$.
The remaining deviation of the period with respect to $T_{0}$ results
from the nonlinear contributions due to the dynamics of the gain and
of the absorber and to phase-amplitude coupling. Finally we note that,
as these temporal LSs are periodic orbits found in the long delay
limit, they can be considered in principle as orbits approaching an
homoclinic solution in the limit $\tau\rightarrow\infty$.

\section{Bifurcation analysis}

\begin{figure}
\centering{}%
\begin{tabular}{ll}
(a) & (b)\tabularnewline
\includegraphics[bb=0bp 0bp 240bp 155bp,clip,width=0.5\columnwidth]{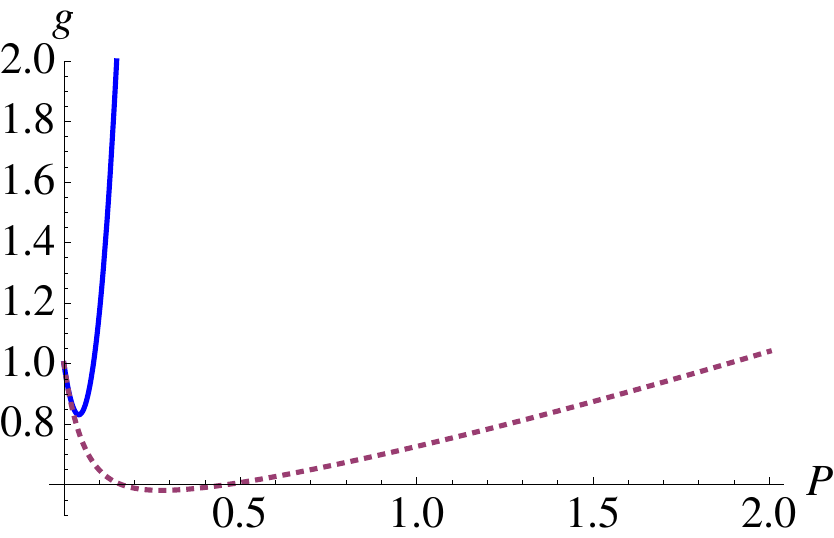} & \includegraphics[bb=0bp 0bp 240bp 155bp,clip,width=0.5\columnwidth]{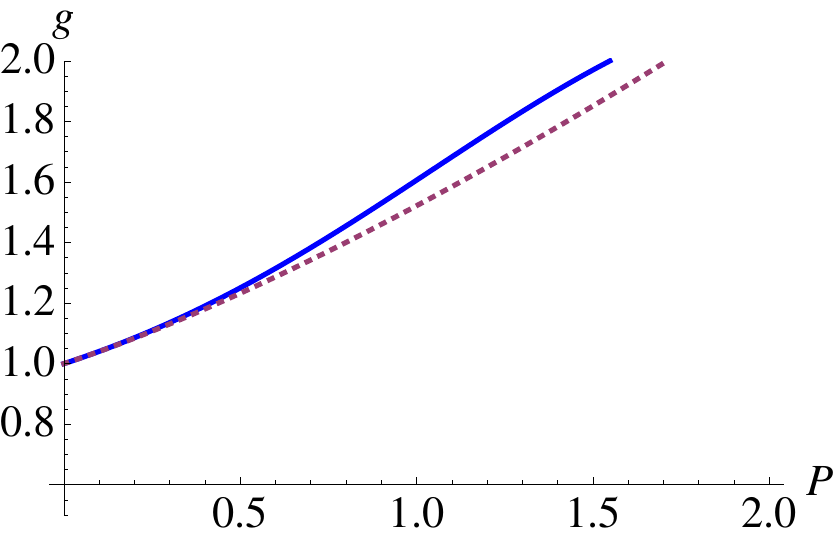}\tabularnewline
\end{tabular}\caption{(Color online) Branch for the single pulse solution as a function
of the normalized gain $g$ for strong (a) and weak (b) absorber nonlinearity.
We represent the pulse energy as given by the weakly nonlinear analysis
assuming a hyperbolic secant in blue, while the red dotted line is
the result of a non-perturbative analysis, assuming a Dirac pulse
shape. Only in the strong absorber regime in (a) $s=30$ and $Q_0=0.3$ 
we find a temporal LS bistable with the off solution while in (b)
$s=5$ and 
$Q_0=0.01$, the pulse develops only above the lasing threshold.
In this case a good agreement between the weakly nonlinear analysis
and the non-perturbative analysis is found. Other parameters are $\alpha=\beta=0$.
\label{fig:New_vs_Haus}}
\end{figure}

\begin{figure}
\centering{}\includegraphics[bb=0bp 0bp 750bp 450bp,clip,width=1\columnwidth]{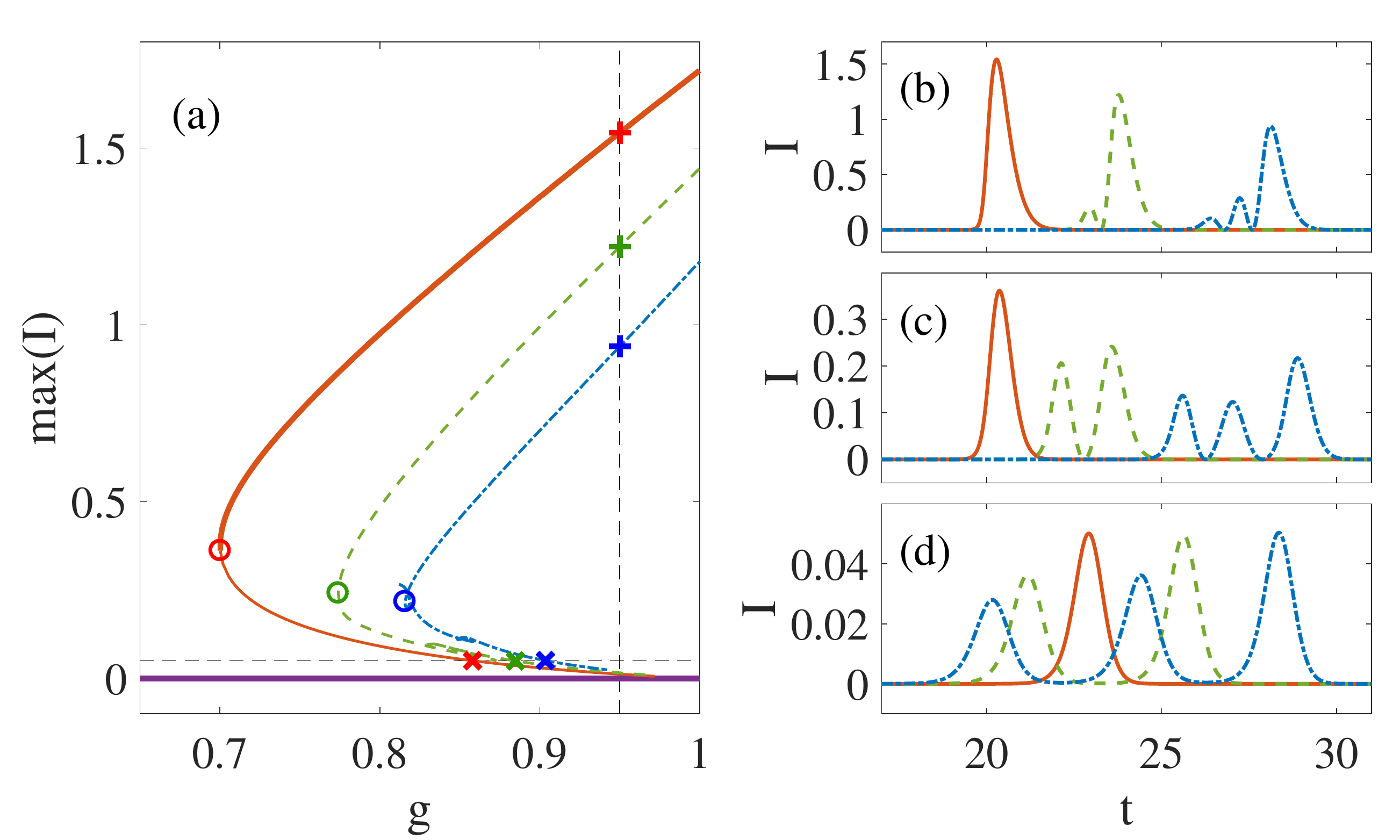}\caption{(Color online) (a) Multiple branches of temporal LSs as a function
of the normalized gain $g$. In addition to the main branch appearing
for the lowest values of the gain at $g_{SNL}^{\left(1\right)}=0.7$,
other branches appear via additional SNL bifurcations at $g_{SNL}^{\left(2\right)}=0.774$,
$g_{SNL}^{\left(3\right)}=0.816$, and consist in pulses composed
of two or three maxima. The temporal profiles for the intensity are
presented in panel (b-d), where, from top to bottom, we show the solutions
on the upper branches (b) at $g=0.95$ (see vertical dashed line),
the profiles at the saddle points (c), and those on the lower branch
(d) with $\mathrm{max}\left(I\right)=0.05$ (see horizontal line).
Other parameters are $\left(\alpha,\beta\right)=\left(0,0\right)$.
\label{fig:2}}
\end{figure}

\paragraph*{The main solution branch}

We start by recalling the main characteristics of our temporal LSs
setting $\alpha=\beta=0$. We operate in a regime of bistability in
which, in addition to the stable off solution, two solutions that
consist in temporal LSs exist. One is unstable and corresponds to
a low intensity temporal pulse while the stable solution is the one
of high intensity. The temporal LSs appear as a saddle-node bifurcation
of limit cycle (SNL) below the lasing threshold, see Fig.~\ref{fig:1}(a),
where we represented the maximal intensity of the pulse while Fig.~\ref{fig:1}(b)
shows the deviation of the solution period $T-T_{0}$. We notice that
the period of the stable portion of the branch is a decreasing function
of $g$. As noted in \cite{JCM-PRL-16,CJM-PRA-16}, this results in
repulsive interactions between temporal LSs as a gain depletion created
by a LS will accelerate the next one away from it. We represent the
temporal profile of the stable LS branch in Fig.~\ref{fig:1}(c),
where the multiscale nature of the solution is apparent. While the
optical pulse length is $\tau_{p}\sim1$, the gain recovery is $3\tau_{g}\sim75$.
The inset Fig.~\ref{fig:1}(e) details the fast component of the
LS. The unstable LS, that plays the role of a separatrix between the
stable LS and the off solution, is represented in Fig.~\ref{fig:1}(d).
We also show in Fig.~\ref{fig:1}(a) the dominant CW solution (the
blue line). We stress that in our regime of localization the CW solutions
are still supercritical and only develop above the lasing threshold.
As such, we do not have bistability for the CW solution.

The typical pulse energy for the upper branch is $P\sim1$, see Fig.~\ref{fig:1}(e),
and $P\sim0.1$ for the lower one, see Fig.~\ref{fig:1}(f). As such
the absorber is operated in a strong saturation regime for which $sP\gg1$.
This regime is far beyond the reach of the usual hyperbolic secant
ansatzes that allow finding values of the pulse energy and of the
pulsewidth. Indeed, these hyperbolic secant ansatzes are correct only
if the absorber saturation can be expanded up to second order, e.g.
$\exp\left(-sP\right)\sim1-sP+\left(sP\right)^{2}/2$. On the contrary,
New's approach of mode-locking \cite{N-JQE-74} only considers infinitely
narrow pulses, e.g., Dirac deltas, but does not necessitate any approximation
on the pulse energy. In our case, this second approach gives a much
better agreement with exact numerics, although the details of the
pulse shape and chirp cannot be obtained. The comparison of both approaches
is depicted in Fig.~\ref{fig:New_vs_Haus} for the regime of strong
and weak nonlinearities. The details of the calculations can be found
in the appendix, for the simple case where $\alpha=\beta=0$. We notice
that only in the strongly nonlinear regime one can obtain a sub-critical
branch and bistability with the off solution. Also, only the beginning
of the lower branch of solution is properly reproduced by the hyperbolic
secant solution, since in this situation the pulse energy can be made
arbitrarily small. While bistability is preserved by both approaches,
neither the upper branch nor the folding point can be properly obtained
using the hyperbolic secant ansatz. New's approach is much more indicative
for the extend of the bistable region and the pulse energy, if one
compares with the results in Fig.~\ref{fig:1} although it does not
allow finding the details nor the possible instabilities of the temporal
LSs. Finally, we note in Fig.~\ref{fig:New_vs_Haus}(b) that for
more standard parameters for PML, i.e., $s=5$ and 
$Q_0=0.01$, the pulsed solutions develops only above the lasing threshold an that
in this case a good agreement between the weakly nonlinear analysis
and the non-perturbative analysis is found. This comparison between
the standard approaches of PML justifies the need for a detailed bifurcation
analysis using path continuation techniques to fully study the localization
regime.

\paragraph*{Multi-peaked solutions }

Still setting $\alpha=\beta=0$, we depict in Fig.~\ref{fig:2}(a)
how, in addition to the main solution branch, additional solutions
appear while increasing the bias current. We only present the first
three branches bifurcating upon increasing $g$, yet additional solutions
continue to appear at an increased rate when $g\rightarrow1$. However,
their evaluation becomes numerically tedious. We represent the temporal
profiles of the intensity at $g=0.95$ on the upper part of the three
branches in Fig.~\ref{fig:2}(b), at their respective folding points
in Fig.~\ref{fig:2}(c), and on the lower part of the branches close
to their appearance threshold, in Fig.~\ref{fig:2}(d). The low intensity
branches are composed of LSs with an increasing number of bumps, similar
to the molecules found for dissipative solitons systems, see e.g.
\cite{GS-LNP-08}. Yet, the dynamics of the gain prevents, with parameters
typical of semiconductors, the creation of stable molecules. As mentioned
earlier, the gain dynamics induces a strong repulsion. All the multi-bump
solutions evolve toward single pulse solutions when they reach the
upper branch at high values of $g$.

\paragraph*{Secondary Andronov-Hopf bifurcation}

\begin{figure}
\begin{centering}
\includegraphics[bb=50bp 0bp 740bp 445bp,clip,width=1\columnwidth]{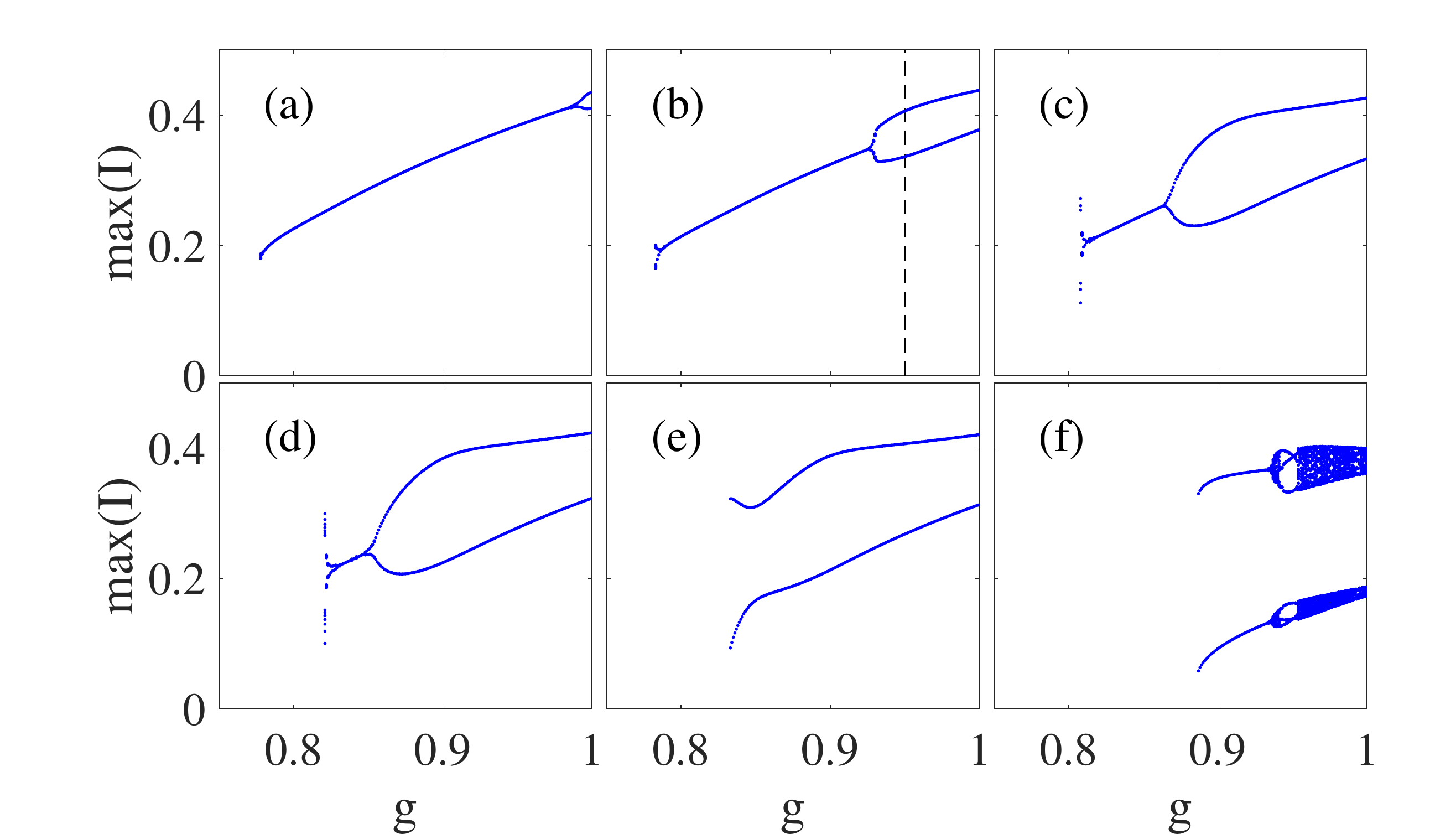}
\par\end{centering}
\centering{}\caption{(Color online) (a-f) Stable branch of temporal LS as a function of
the normalized gain $g$, obtained via direct numerical simulation
of Eqs.~(\ref{eq:VT1}-\ref{eq:VT3}). The dynamics are represented
by the extrema of the time traces of the maximal pulse intensities.
(a) $\alpha=3.7$, most of the branch is stable, although a secondary
supercritical AH bifurcation is obtained for $g_{H_{1}}\simeq0.98$.
(b) For $\alpha=3.8$, the quasi-periodic regime shifts toward lower
$g_{H_{1}}\simeq0.92$ while another, subcritical, AH occurs at lower
current $g_{H_{2}}\simeq0.81$. (c) At $\alpha=4$ and (d) $\alpha=4.05$,
the two quasi-periodic solutions come closer and finally merge at
$\alpha=4.1$ as shown in (e). Higher values of $\alpha=5$ lead to
a quasi-periodic cascade (f). Other parameters are $\beta=0.5$. \label{fig:3}}
\end{figure}

\begin{figure}
\begin{centering}
\includegraphics[clip,width=1\columnwidth]{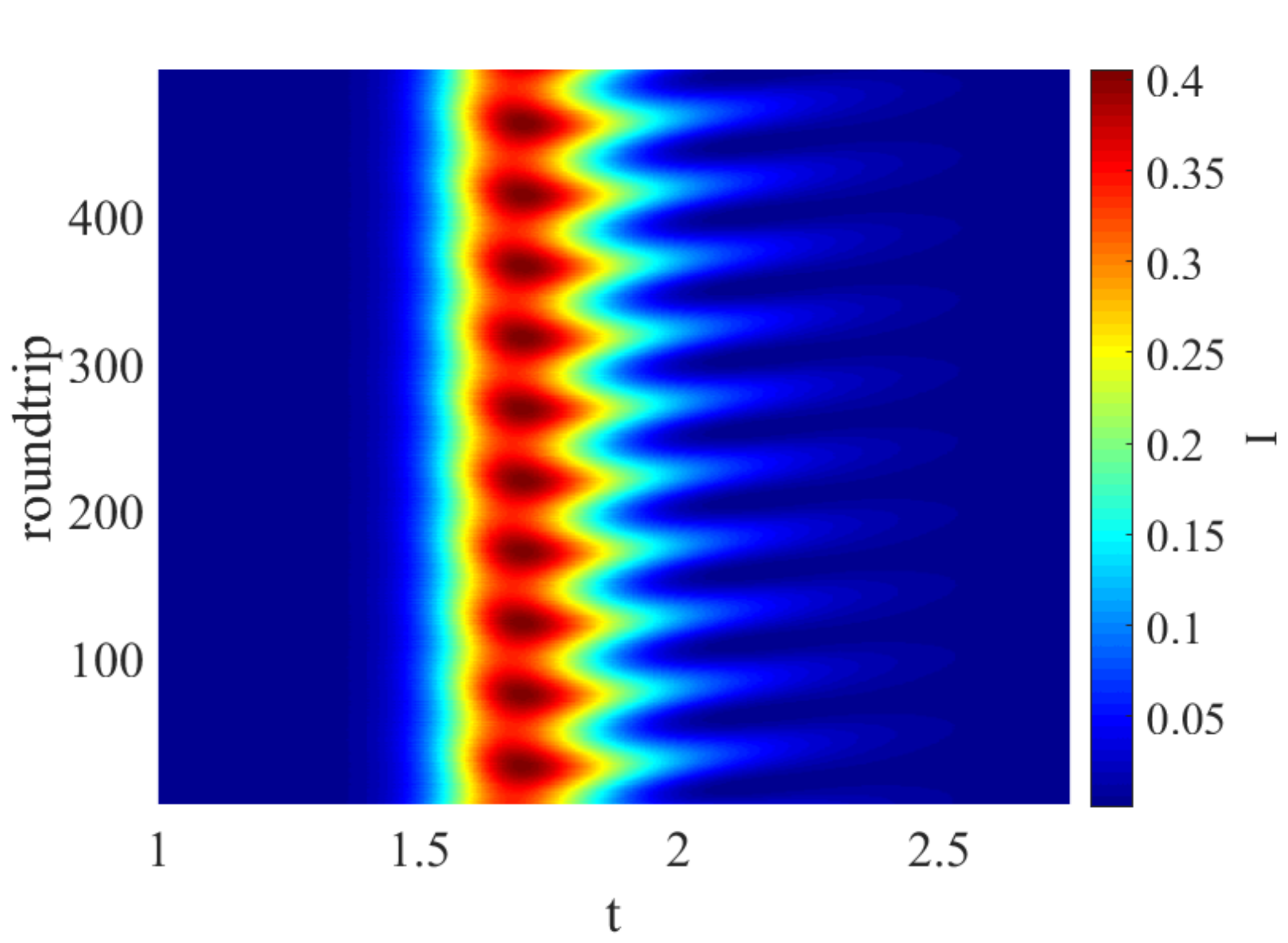}
\par\end{centering}
\centering{}\caption{(Color online) Quasi-periodic limit cycle time trace obtained with
$g=0.95$, see dotted line in Fig.~\ref{fig:3}(b), using a space-time
representation. Other parameters are $\alpha=3.8$ and $\beta=0.5$.
\label{fig:space_time}}
\end{figure}

We now turn our attention toward the dynamics found for large, yet
realistic, values of the linewidth enhancement factors in the gain
and the absorber sections. For the gain, we set $\alpha=3.7$ while
for the absorber we set $\beta=0.5$. As the latter is operated below
the transparency, the effects of band filling are much weaker, which
justifies using a much smaller value of the Henry factor. As the bifurcation
study of quasi-periodic orbits is not currently possible with ddebiftool,
we performed direct numerical simulations of Eqs.~(\ref{eq:VT1}-\ref{eq:VT3}).
We integrated Eqs.~(\ref{eq:VT1}-\ref{eq:VT3}) with a fourth order
Runge-Kutta with Hermite interpolation of the time-delayed term and
a step size $\Delta t=10^{-2}$. We depict in Fig.~\ref{fig:3}(a)
the bifurcation diagram obtained by direct numerical integration,
performing a parameter sweep in $g$, upward an downward starting
from a central value. Using numerical integration, we can only show
the upper part of the main branch, as it is the only stable solution.
We observe that the main solution branch, that actually consists in
a strongly nonlinear (pulsating) limit cycle, develops a secondary
oscillation frequency (typically ranging between a few tens and a
few hundreds of round-trips) when the gain is increased toward the
lasing threshold. This slowly evolving orbit during which the pulse
parameters are oscillating in time is depicted in Fig.~\ref{fig:space_time}
using a space-time representation for $\alpha=3.8$. Here, we show
the evolution of the pulse train, from one round-trip to the next.
This diagram allows us to identify this secondary Andronov-Hopf (AH)
instability as a trailing edge instability. As it occurs for large
values of $\alpha$ and increasing values of the gain, we posit it
is a dispersive (phase) instability. 

The evolution of this emerging limit cycle is depicted in Fig.~\ref{fig:3}(c)
for higher values of $\alpha=4$ which shifts the secondary AH to
lower values of $g$ while another subcritical AH appears at a lower
value of the gain. In this regime, the region of stable operation
is delimited by these two AH bifurcations. Using higher values of
$\alpha=4.1$ leads to a collision and a merging of these two quasi-periodic
solutions, see Fig.~\ref{fig:3}(d,e). In this regime, stable LSs
do not exist and solely oscillating quasi-periodic solutions are found.
For larger values of $\alpha$ and high gain, a typical transition
to irregular dynamics via quasi-periodicity is observed, which is
visible in Fig.~\ref{fig:3}(f) where the maximal pulse intensity
shows quasi-continuous values. 

\begin{figure}
\begin{centering}
\includegraphics[clip,height=4.4cm]{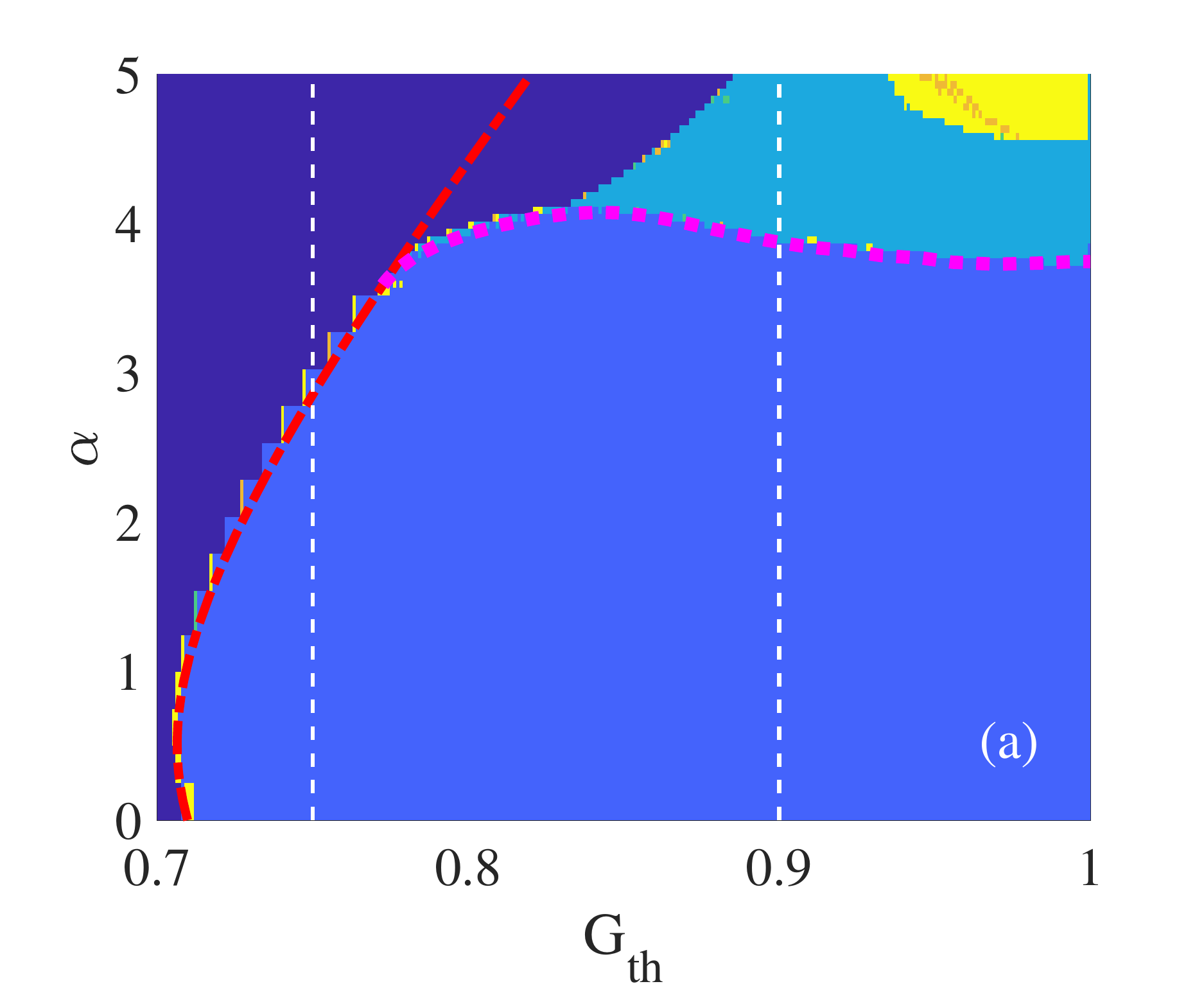}\includegraphics[bb=0bp 0bp 375bp 445bp,clip,height=4.4cm]{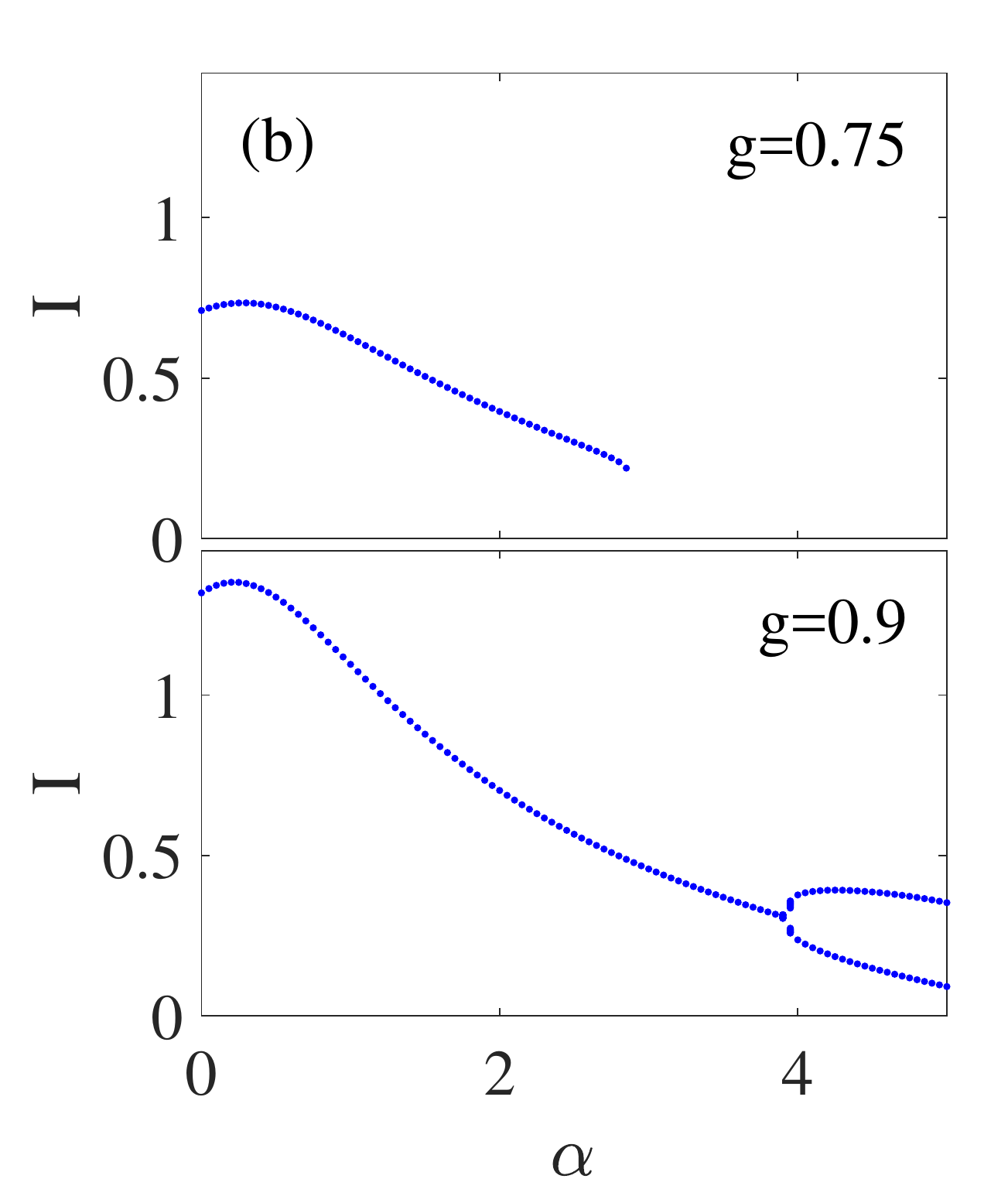}
\par\end{centering}
\centering{}\caption{(Color online) (a) Two dimensional bifurcation diagram as a function
of $g$ and $\alpha$. Dark blue: off solution. Blue: stable LS,
light blue: periodically oscillating. Yellow: quasi-periodic LS oscillation.
(b) Vertical cuts of the diagram in (a) obtained for $g_{1}=0.75$
and $g_{2}=0.9$ as a function of $\alpha$. One notices for low values
of $\alpha$ an extended domain of stability ranging from the appearance
of the SNL bifurcation for low $g$ toward threshold $g=1$. We superposed
the results obtained with ddebiftool where the red dash-dotted and
pink dashed lines correspond to the SNL and the secondary AH, respectively.
Other parameters are $\beta=0.5$. \label{fig:AH_diag}}
\end{figure}

In order to understand how the various regimes are connected together,
we performed a double scan in the parameters $\alpha$ and $g$. Our
results are summarized in Fig.~\ref{fig:AH_diag}. We superposed
to these numerical results the evolution of the SNL point for the
primary branch as well as the secondary AH point found by using ddebiftool,
finding a good agreement. While ddebiftool cannot track the emerging
solution, it can identifies the secondary AH point, which is actually
a Neimark-Sacker bifurcation. First, we note in Fig.~\ref{fig:AH_diag}(a)
that the SNL values depend rather weakly on $\alpha$ and that the
minimal value of $g_{SNL}$ is not attained for $\alpha=0$. This
is due to the presence of a non-zero value of $\beta$ and a small
value of $\alpha$ can compensate for the chirp created by the absorber.
However, the Lorentzian filter in Eq.~(\ref{eq:VT1}) limits the
optical bandwidth of the field and high values of $\alpha$ induce
additional chirp for the pulses which, in turn, creates additional
optical bandwidth that gets absorbed by the filter. As such, highly
chirped pulses experience more losses and can not exist for too low
values of the gain, which explains why the SNL point increases in
$g$ for large values of $\alpha$. Also, one notices a different
scenario depending on the value of $\alpha$. For low values of $\alpha$
an extended domain of stability ranges from the appearance of the
SNL bifurcation for low $g=g_{SNL}$ toward threshold $g=1$. For
higher values of $\alpha\in\left[3.7,3.75\right]$, the solution stability
is still governed by the SNL for low values of $g$ but by the AH
bifurcation that is crossed at higher values of $g$. We notice that
the two AH bifurcations depicted in Fig.~\ref{fig:3}(b-d) are actually
stemming from the same AH curve in the $\left(g,\alpha\right)$ plane
that can be crossed twice upon increasing $g$. For higher values
of $\alpha\in\left[3.75,4.1\right]$, the stable domain for the LS
is enclosed between the two AH points. For values of $\alpha>4.1$,
where the two AH points merged, the only kind of LS that exists is
an oscillating one.

Similar diagrams were obtained for other values of parameters, and
we note that, while it is not the case with $\beta=0.5$, some bistability
between steady and oscillating solutions could be observed in a finite
interval of $\left(g,\alpha\right)$ by setting $\beta=0$. It stands
to reason that this bistability could be preserved for low values
of $\beta$ and adapted values of the other parameters as, e.g., $\left(\kappa,q_{0},s\right)$.

\begin{figure*}
\begin{centering}
\includegraphics[clip,width=2\columnwidth]{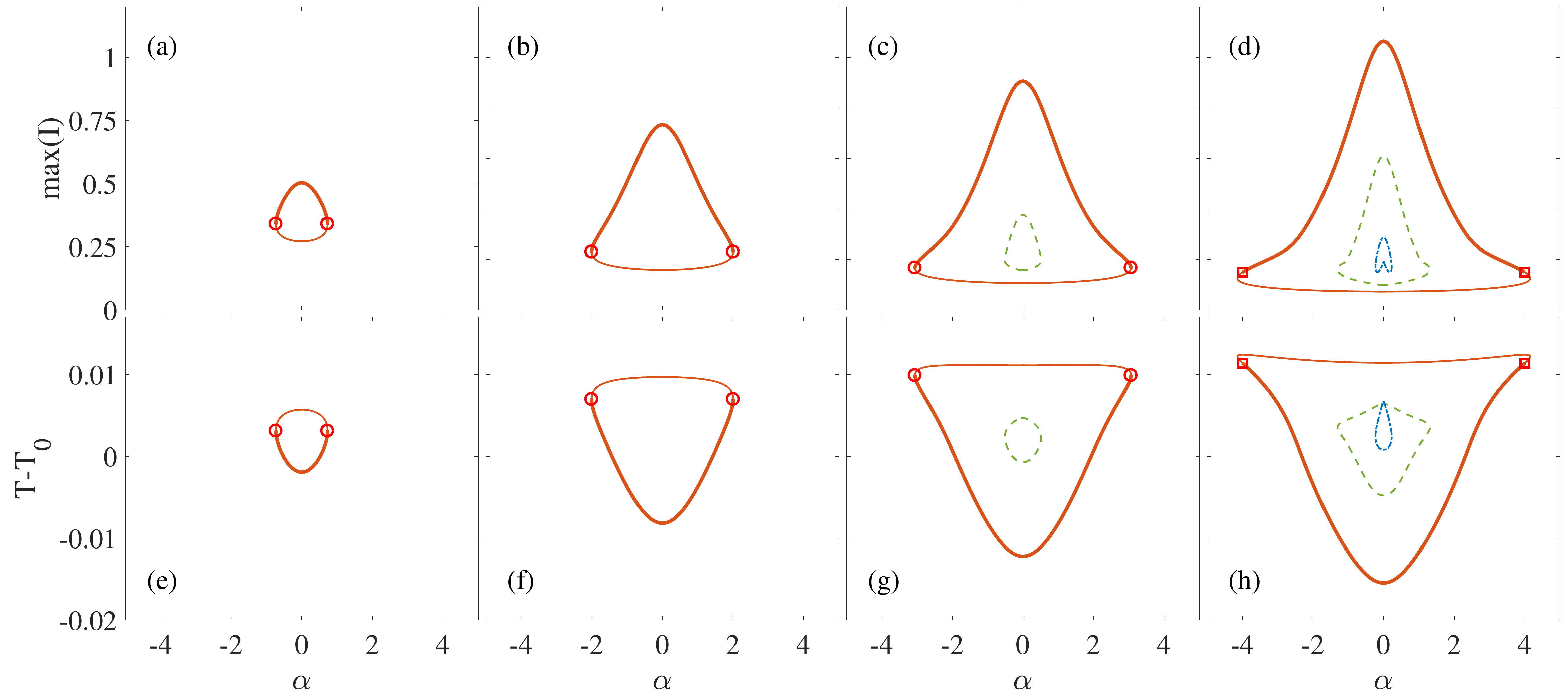}
\par\end{centering}
\centering{}\caption{(Color online) (a) Two-dimensional bifurcation diagram in $\alpha$
for increasing values of $g$. We represent the maximum intensity
(a-d) and the period deviation (e-h). The value of the gain is $g=0.707$
(a,e), $g=0.745$ (b,f), $g=0.784$ (c,g) and $g=0.822$ (d,h). Stability
is indicated with thick lines. Additional solution loops are born
via saddle-node bifurcations upon increasing $g$. Other parameters
are $\beta=0$. \label{fig:SimpleLoops}}
\end{figure*}

\paragraph*{Organization of solutions in the $\left(g,\alpha\right)$ plane}

We now turn our attention to how the multiple solution branches depicted
in Fig.~\ref{fig:2} organize by making a three-dimensional bifurcation
diagram of the LS solutions where our control parameters are $\left(\alpha,g\right)$.
First, we set the linewidth enhancement factor of the absorber to
$\beta=0$. Our results are summarized in Fig.~\ref{fig:SimpleLoops},
where we present various slices of the diagram, the solution curves
in $\alpha$, for increasing values of $g$. We represent the maximum
pulse intensity as well as the period deviation of the solution. As
we want to emphasize the solution structure, we extended our analysis
to negative values of $\alpha$. For $\beta=0$, the diagram is perfectly
symmetrical, since negative values of $\alpha$ simply consist in
taking the complex conjugate of Eq.~(\ref{eq:VT1}). Firstly, we
note in Fig.~\ref{fig:SimpleLoops}(a,e) that the solution loop folds
for larger values of $\alpha$, here $\alpha_{fold}\sim\pm1$. As
previously mentioned, $\alpha$ induces additional chirp of the solution
which limits the region of existence of the LSs. A higher value of
$g$ depicted in Fig.~\ref{fig:SimpleLoops}(b,f) allows the LS solution
to exists at higher values of $\alpha$. This evolution of the folding
point is another representation of the evolution of the SNL curve
shown in Fig.~\ref{fig:AH_diag}. One notices that the solution structure,
at low $g$, resembles a paraboloid growing in radius when $g$ is
increased, that then deforms nonlinearly. At higher values of $g$,
an additional solution loop emerges, see Fig.~\ref{fig:SimpleLoops}(c,g).
This loop corresponds to the solutions with a double pulse, and it
grows in radius at higher $g$, see Fig.~\ref{fig:SimpleLoops}(d,g),
where also a third loop with a three peaked solution emerges.

\begin{figure}
\begin{centering}
\includegraphics[bb=0bp 0bp 600bp 600bp,clip,width=1\columnwidth]{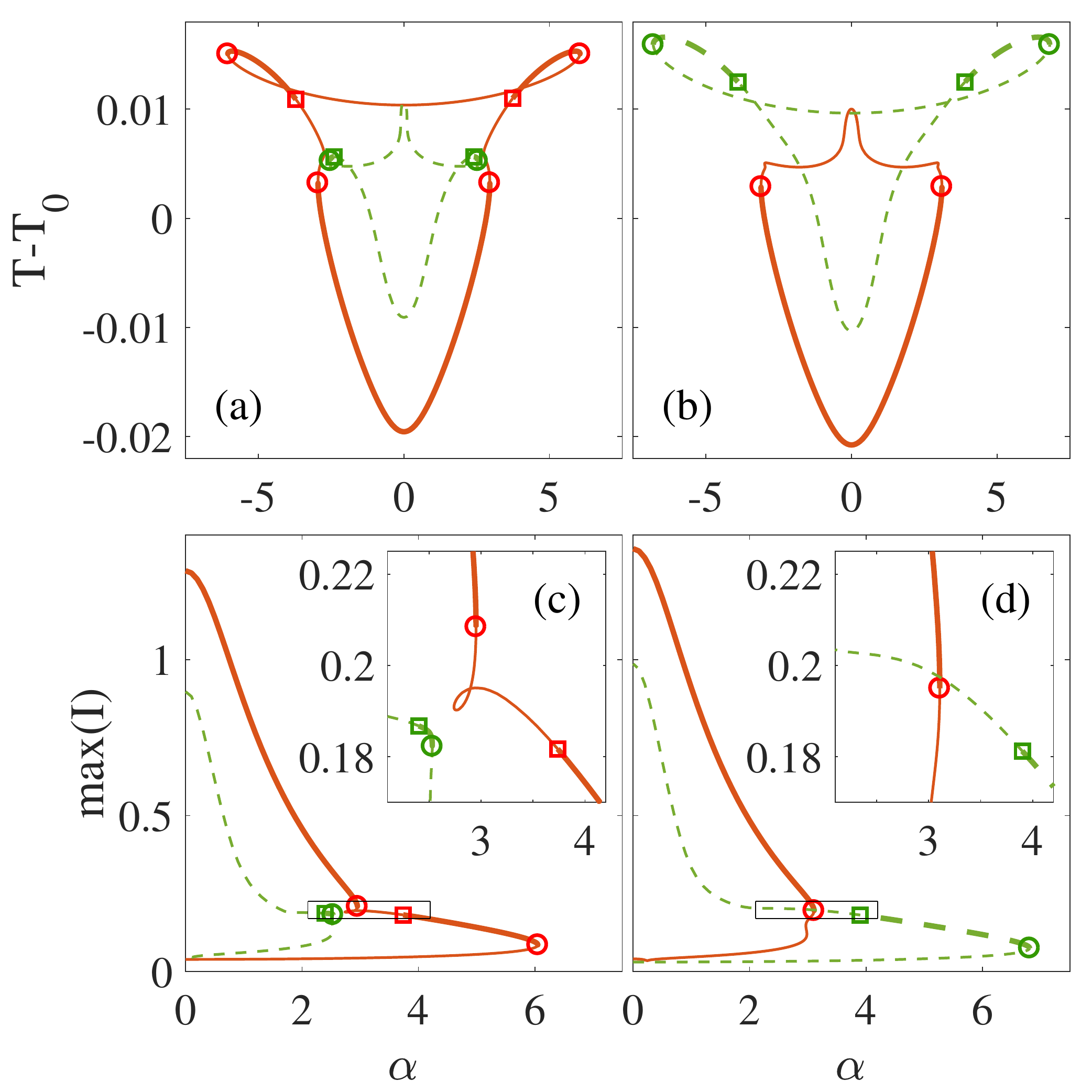}
\par\end{centering}
\centering{}\caption{(Color online) (a-d) Bifurcation diagrams as a function of $\alpha$
for two close-by values of the gain, $g=0.879$ (a,c) and $g=0.898$
(b,d). We represent in (a,b) the deviation of the period and the maximum
intensity in (c,d). Stable solutions are depicted in thick lines.
Open circles denote the fold positions whereas squares indicate the
AH bifurcation points. Other parameters are $\beta=0$. \label{fig:transLoops}}
\end{figure}

\begin{figure}
\centering{}\includegraphics[bb=0bp 0bp 458bp 295bp,clip,width=1\columnwidth]{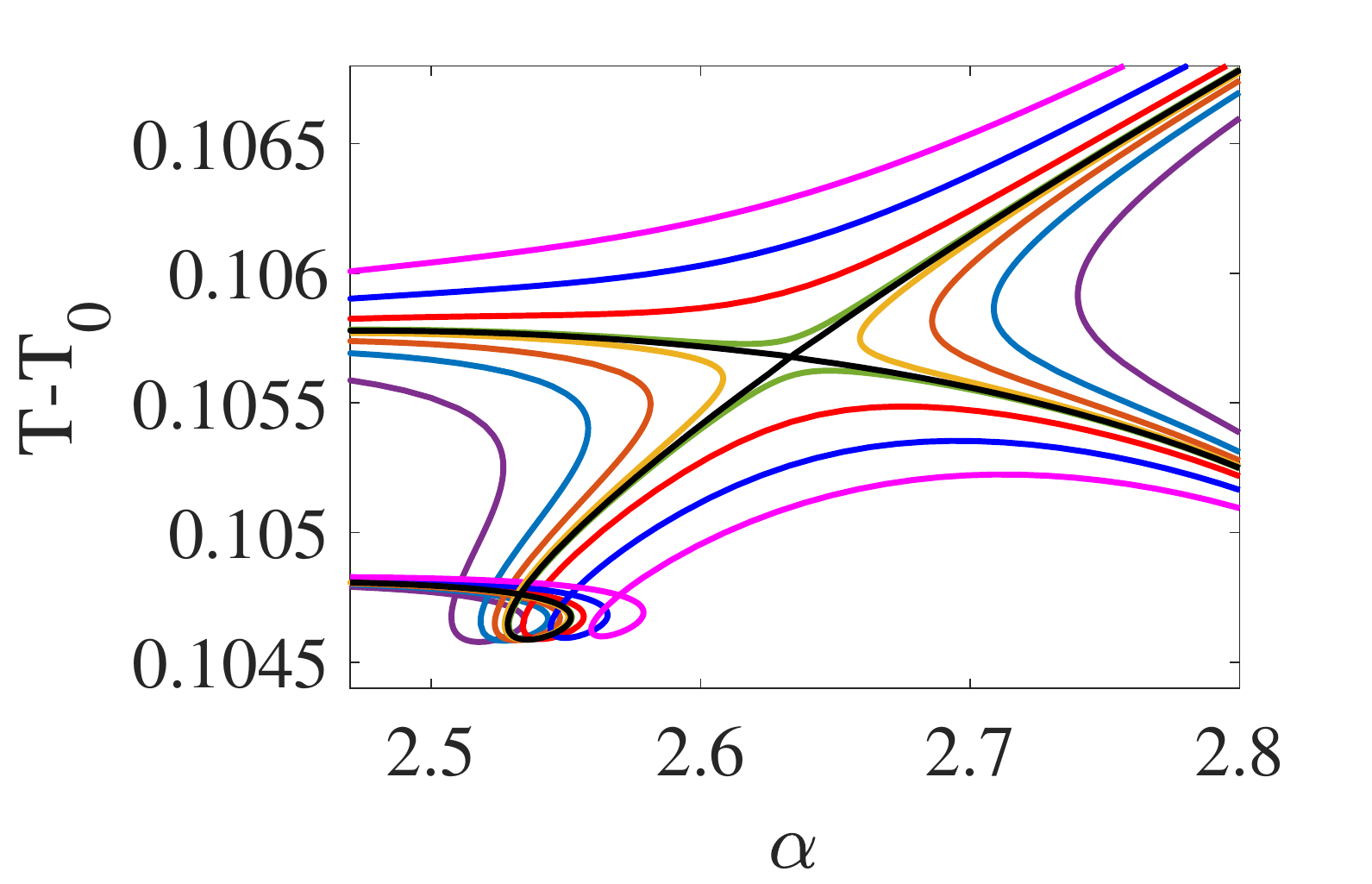}\caption{(Color online) (a) Zoom around the transcritical bifurcation depicted
in Fig.~\ref{fig:transLoops}. The deviation of the period is shown.
The values of $g$ are interspersed in the interval $g\in\left[0.8793;0.8812\right]$.
Other parameters are $\beta=0$. \label{fig:transZoom}}
\end{figure}

We depict the interaction occurring between these various solution
loops when they become of comparable amplitude. The interactions between
the primary and the secondary loops is described in Fig.~\ref{fig:transLoops}.
For $g=0.879$, the outer branch, that is the stable solution for
large $\alpha$, develops a pair of folds via a cusp bifurcation.
This cusp takes the form of an additional loop along the branch, if
one represents the maximum pulse intensity, see the inset in Fig.~\ref{fig:transLoops}(c).
For $g=0.898$, the primary and secondary solution loops have crossed
each other via a transcritical bifurcation. This mechanism is important
because, at high values $\alpha$, it is now the secondary branch,
initially unstable and showing solutions with two peaks, that is responsible
of giving the stable solution with a single peak, see Fig.~\ref{fig:transLoops}(b)
and the inset in Fig.~\ref{fig:transLoops}(d). As previously mentioned,
the mechanism by which the two solution loops can cross is a transcritical
bifurcation. We depict this mechanism by which the solution curves
are allowed to cross each other in a small vicinity of the bifurcation
point in Fig.~\ref{fig:transZoom}.

\begin{figure*}
\begin{centering}
\includegraphics[bb=0bp 0bp 1350bp 600bp,clip,width=2\columnwidth]{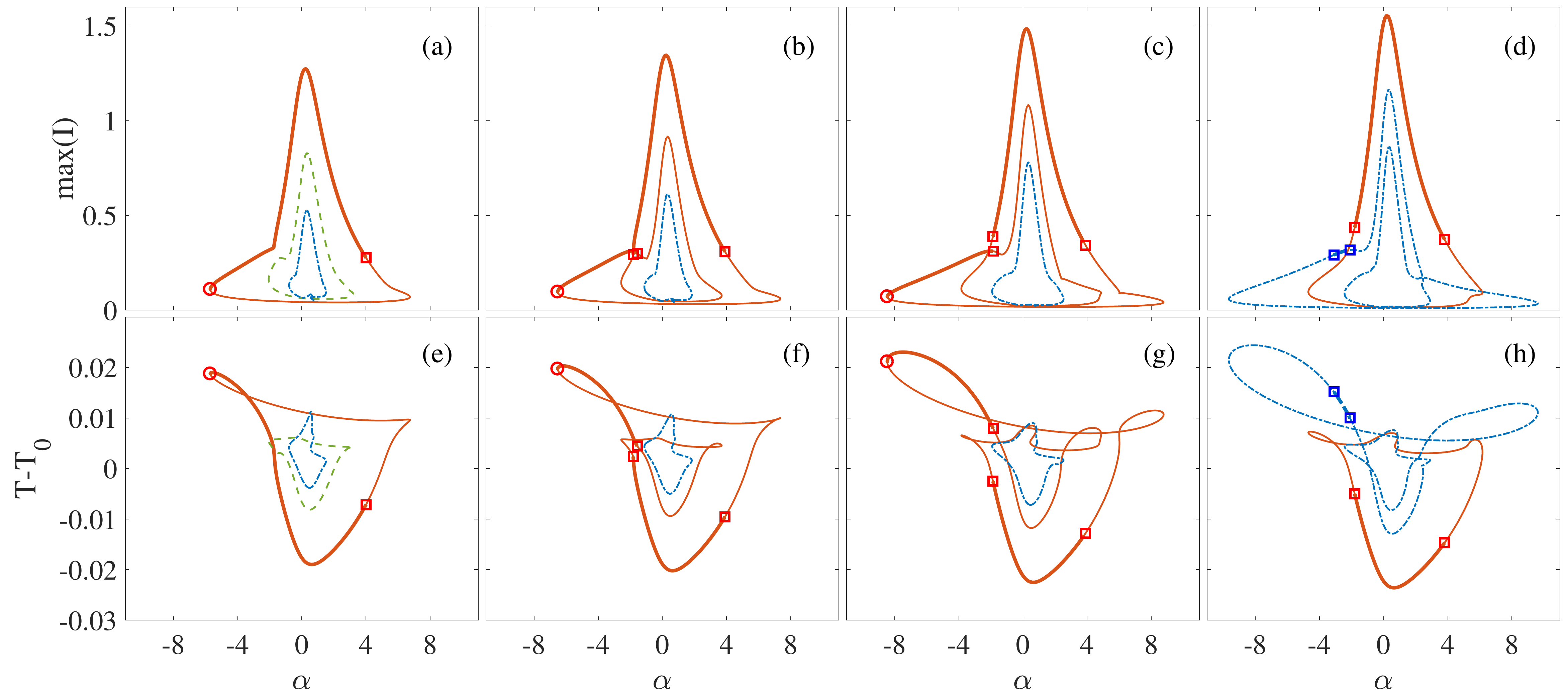}
\par\end{centering}
\centering{}\caption{(Color online) (a) Two-dimensional bifurcation diagram in $\alpha$
for increasing values of $g$. We represent the maximum intensity
(a-d) and the period deviation (e-h). The value of the gain is $g=0.879$
(a,e), $g=0.898$ (b,f), $g=0.937$ (c,g) and $g=0.956$ (d,h). Stability
is indicated with thick lines. Open circles denote the fold positions
whereas squares indicate the AH bifurcation points. The different
branches intersect asymmetrically when increasing $g$ at $\beta=0.5$.
\label{fig:ComplexLoops}}
\end{figure*}

Finally, we consider how to this bifurcation scenario changes when
$\beta\neq0$. We set $\beta=0.5$ and the first consequence of having
$\beta\neq0$ is that the symmetry $\alpha\rightarrow-\alpha$ of
the bifurcation diagrams depicted in Fig.~\ref{fig:SimpleLoops}
is broken. While for $\beta=0$ pairs of transcritical bifurcations
would appear symmetrically and reconnect parts of some solution loops
with others on both sides, it is not the case anymore. Our results
are depicted in Fig.~\ref{fig:ComplexLoops} where we can appreciate
the changes in the bifurcation scenario. While the gradual appearance
of additional solutions is preserved when increasing $g$, see Fig.~\ref{fig:ComplexLoops}(a,e),
we notice that the transcritical bifurcations appear in an alternated
way, first for a negative value of $\alpha$, see Fig.~\ref{fig:ComplexLoops}(b,f),
then for a positive, yet different value of $\alpha$, see Fig.~\ref{fig:ComplexLoops}(d,h).
This has the consequence of giving the solution surface the visual
\emph{appearance} of a Klein bottle, as depicted for instance in Fig.~\ref{fig:ComplexLoops}(c,g).
Here, the apparent self-intersection of the primary solution branch
is visible. However, while the branch seems to self-intersect when
looking at the maximum pulse intensity, another measure of the branch
would give a different representation.

\section{The exponential Haus master equation }

In this section, we turn our attention toward the predictions given
by a different approach that is based upon a partial differential
equation (PDE) instead of a DDE. This modified Haus master equation
considers the evolution of a pulse on a slow time scale that corresponds
to the number of round-trips in the cavity. As such, this iterative
pulse mapping can be much more efficient computationally. In addition,
while the LSs are periodic solutions of a DDE, they become steady
states of a one-dimensional PDE, which can lead to further bifurcation
analysis. For instance, the branches of periodic solutions of a PDE
can be computed using the path continuation methods while the quasi-periodic
solutions of the DDE cannot be evaluated with ddebiftool at the moment.
Another argument that makes the PDE approach appealing. One can actually
restrict the numerical domain along the propagation axis to a box
that is a few times the extension of the optical pulse. That way,
the long gain recovery, during which the field is zero can be discarded,
which results in a much reduced number of degree of freedom during
the continuation. 

We outline how the DDE given by Eqs.~(\ref{eq:VT1}-\ref{eq:VT3})
can be recast into a PDE. We have seen that, at the lasing threshold,
the maximum gain mode needs to have a frequency shift $\omega_{0}=\Theta/\tau$.
While the frequency shift is arbitrarily small in the long delay limit,
the phase shift per pass $\Theta$ is not. It is essential, as it
compensates for the index variation created by the active medium after
one round-trip. Within the framework of an iterative pulse model as
the Haus master equation, that does not contain anymore proper boundary
conditions for the field, this frequency shift has to be canceled
out before making the correspondence between the DDE and the PDE.
We perform the change of variable $A\left(t\right)=E\left(t\right)e^{-i\omega_{0}t}$
in order to cancel this rotation, which leads to the modified field
equation 
\begin{eqnarray}
\frac{\dot{E}}{\gamma}-i\frac{\Theta}{\gamma\tau}E & = & R_{\tau}e^{i\Theta}E_{\tau}-E,\label{eq:VTN1}
\end{eqnarray}
while the carrier equations are identical simply setting $A\rightarrow E$.
Following the method depicted in \cite{CJM-PRA-16}, the Eqs.~(\ref{eq:VTN1},\ref{eq:VT2},\ref{eq:VT3})
can be transformed into a PDE, taking advantage of the long cavity
limit at which we operate this system experimentally. We do not repeat
the procedure that can be found in \cite{CJM-PRA-16} and only sketch
the reasoning. We start by defining a smallness parameter as the inverse
of the filter bandwidth setting $\varepsilon=1/\gamma$. Physical
intuition dictates that the pulse-width scales as the inverse of the
filter bandwidth and that it is proportional to $\gamma^{-1}=\varepsilon$.
This intuition is confirmed by the numerical continuation. In a related
way, one can foresee that the period of the pulse train scales as
$T_{0}\sim\tau+\gamma^{-1}$, i.e., the period is always larger than
the delay due to causality and the finite response time of the filtering
element that limit the optical bandwidth available. As such, we assume
that the solution is composed of two time scales and write 
\begin{eqnarray}
\frac{d}{dt} & \rightarrow & \frac{\partial}{\partial z}+\varepsilon^{2}\frac{\partial}{\partial s}
\end{eqnarray}
with $\left(z\right)$ governing the fast evolution along the cavity
axis and $s$ depicting the slow dynamics after each round-trip. Following
\cite{GP-PRL-96}, we express the delayed term as 
\begin{eqnarray}
E\left(t+\tau\right) & = & E\left(z+\varepsilon\upsilon,s+\varepsilon^{2}\tau\right),
\end{eqnarray}
which means that the solution after one round-trip is slowly evolving
and drifting. Upon expanding all contributions up to $\mathcal{O}\left(\varepsilon^{3}\right)$,
one finds that the drift term can be canceled setting $\upsilon=-1$.
In other words, the solution at the next round-trip is shifted to
the right, which precisely corresponds to a period of $T_{0}=\tau+\gamma^{-1}$.
Finally, defining a time scale normalized by the round-trip as $\sigma=s/\varepsilon^{2}\tau$
and setting $I=\left|E\right|^{2}$ we find
\begin{eqnarray}
\frac{\partial E}{\partial\sigma}-\frac{1}{2\gamma^{2}}\frac{\partial^{2}E}{\partial z^{2}} & = & \\ \nonumber
    &&\left(\sqrt{\kappa}e^{\frac{1-i\alpha}{2}G-\frac{1-i\beta}{2}Q+i\Theta}-1 +i\frac{\Theta}{\gamma\tau}\right)E,\label{eq:H1a}\\
\frac{\partial G}{\partial z}+\frac{1}{\tau}\frac{\partial G}{\partial\sigma} & = & \Gamma\left(G_{0}-G\right)-e^{-Q}\left(e^{G}-1\right)I,\label{eq:H2a}\\
\frac{\partial Q}{\partial z}+\frac{1}{\tau}\frac{\partial Q}{\partial\sigma} & = & Q_{0}-Q-s\left(1-e^{-Q}\right)I.\label{eq:VT3a}
\end{eqnarray}
We can now invoke the long delay limit and discard in Eq.~(\ref{eq:H1a}-\ref{eq:VT3a}),
all the contributions that are proportional to $1/\tau$. Note that
while the contribution $\Theta/\left(\gamma\tau\right)$ is irrelevant,
we must keep the term $\exp\left(i\Theta\right)$ in Eq.~(\ref{eq:H1a}).
Hence, we obtain the following PDE system
\begin{eqnarray}
\frac{\partial E}{\partial\sigma}-\frac{1}{2\gamma^{2}}\frac{\partial^{2}E}{\partial z^{2}} & = &\left(\sqrt{\kappa}e^{\frac{1-i\alpha}{2}G-\frac{1-i\beta}{2}Q+i\Theta}-1\right)E,\label{eq:H1b}\\
\frac{\partial G}{\partial z} & = & \Gamma\left(G_{0}-G\right)-e^{-Q}\left(e^{G}-1\right)I,\label{eq:H2b}\\
\frac{\partial Q}{\partial z} & = & Q_{0}-Q-s\left(1-e^{-Q}\right)I.\label{eq:H3b}
\end{eqnarray}

The Eqs.~(\ref{eq:H1b}-\ref{eq:H3b}) can be understood as a generalization
of the Haus master equation to large gain and absorption per pass.
Indeed, one of the main advantages of the model of \cite{VT-PRA-05}
is the consideration of large gain and absorption per round-trip,
a feature that is still preserved by the exponential terms in Eqs.~(\ref{eq:H1b}-\ref{eq:H3b}).
The longitudinal variable $\left(z\right)$ identifies as a fast time
variable and represents the longitudinal evolution of the field within
the round-trip. From the inspection of Eqs.~(\ref{eq:H2a},\ref{eq:VT3a})
one can clearly see that the parity symmetry $\left(z\right)\rightarrow\left(-z\right)$,
is being broken by the carrier dynamics that is only first order in
$\partial_{z}$, a symmetry is only recovered upon making the adiabatic
elimination of $G$ and $Q$. Notice that while the regime of a fast
absorber is a meaningful limit, the gain is the slowest variable and
it cannot be eliminated by taking the long cavity limit.

\section{Bifurcation Analysis of the Exponential Haus Equation}

In this section we present the bifurcation analysis of the generalized
Haus master equation described by Eqs.~(\ref{eq:H1b}-\ref{eq:H3b})
and discuss how it is related to that of the DDE model given by Eqs.~(\ref{eq:VT1}-\ref{eq:VT3}).
The temporal LS solutions of Eqs.~(\ref{eq:H1b}-\ref{eq:H3b}) are
slowly drifting oscillating solutions that can be found as steady
states of Eqs.~(\ref{eq:H1b}-\ref{eq:H3b}) by setting 
\begin{eqnarray}
E\left(z,\sigma\right) & = & E\left(z-\upsilon\sigma,\sigma\right)\exp\left(-i\omega\sigma\right)
\end{eqnarray}
which adds a contribution $\left(\upsilon\partial_{z}+i\omega\right)E$
to the right hand side of Eq.~(\ref{eq:H1b}). We recall that the
steady states of Eqs.~(\ref{eq:H1b}-\ref{eq:H3b}) are actually
the periodic solutions of Eqs.~(\ref{eq:VT1}-\ref{eq:VT3}). We
followed the LS solutions of Eqs.~(\ref{eq:H1b},\ref{eq:H3b}) in
parameter space, by using pseudo-arclength continuation within the
pde2path framework \cite{uecker2014}.

\begin{figure}
\begin{centering}
\includegraphics[bb=0bp 0bp 600bp 600bp,clip,width=1\columnwidth]{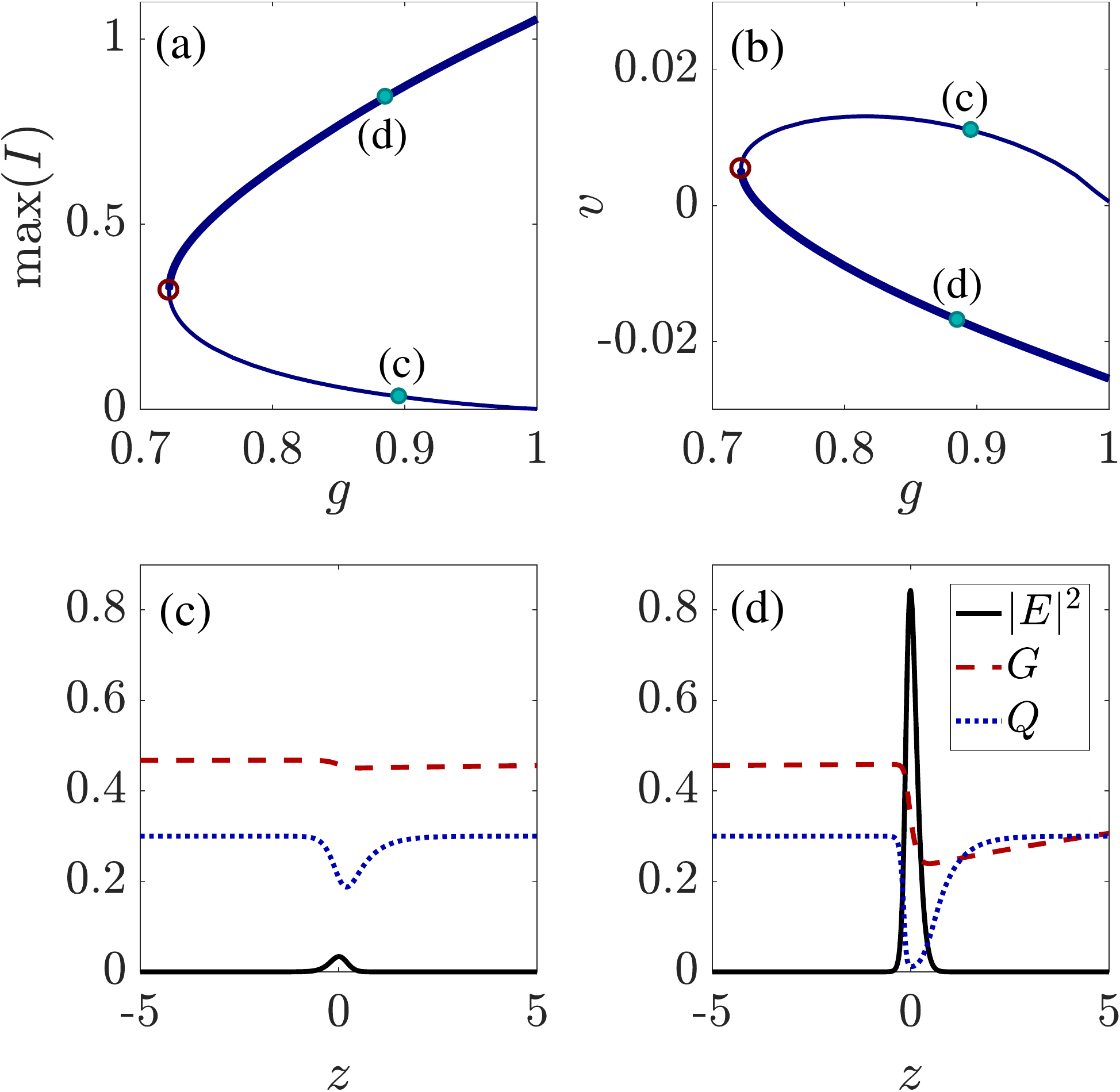}
\par\end{centering}
\centering{}\caption{(Color online) (a,b) Branch of the single temporal LS as a function
of the normalized gain $g$ calculated for $(\alpha,\text{\,\ensuremath{\beta}})=(1.5,\,0.5)$.
We represent (a) the maximum intensity and (b) the drift velocity
$\upsilon$. The LS is stable beyond the saddle-node bifurcation point
$g_{SN}=0.721$ (red circle). (c,d) Two exemplary stationary LS profiles
for the unstable branch (c) and the stable one (d) for $g=0.896$
and $g=0.886$, respectively. Other parameters are $(\gamma,\,\kappa,\,\Gamma,\,Q_{0},\,s)=\left(10,\,0.8,\,0.04,\,0.3,\,30\right)$.
\label{fig:Haus1}}
\end{figure}

\begin{figure}
\begin{centering}
\includegraphics[bb=0bp 0bp 600bp 600bp,clip,width=1\columnwidth]{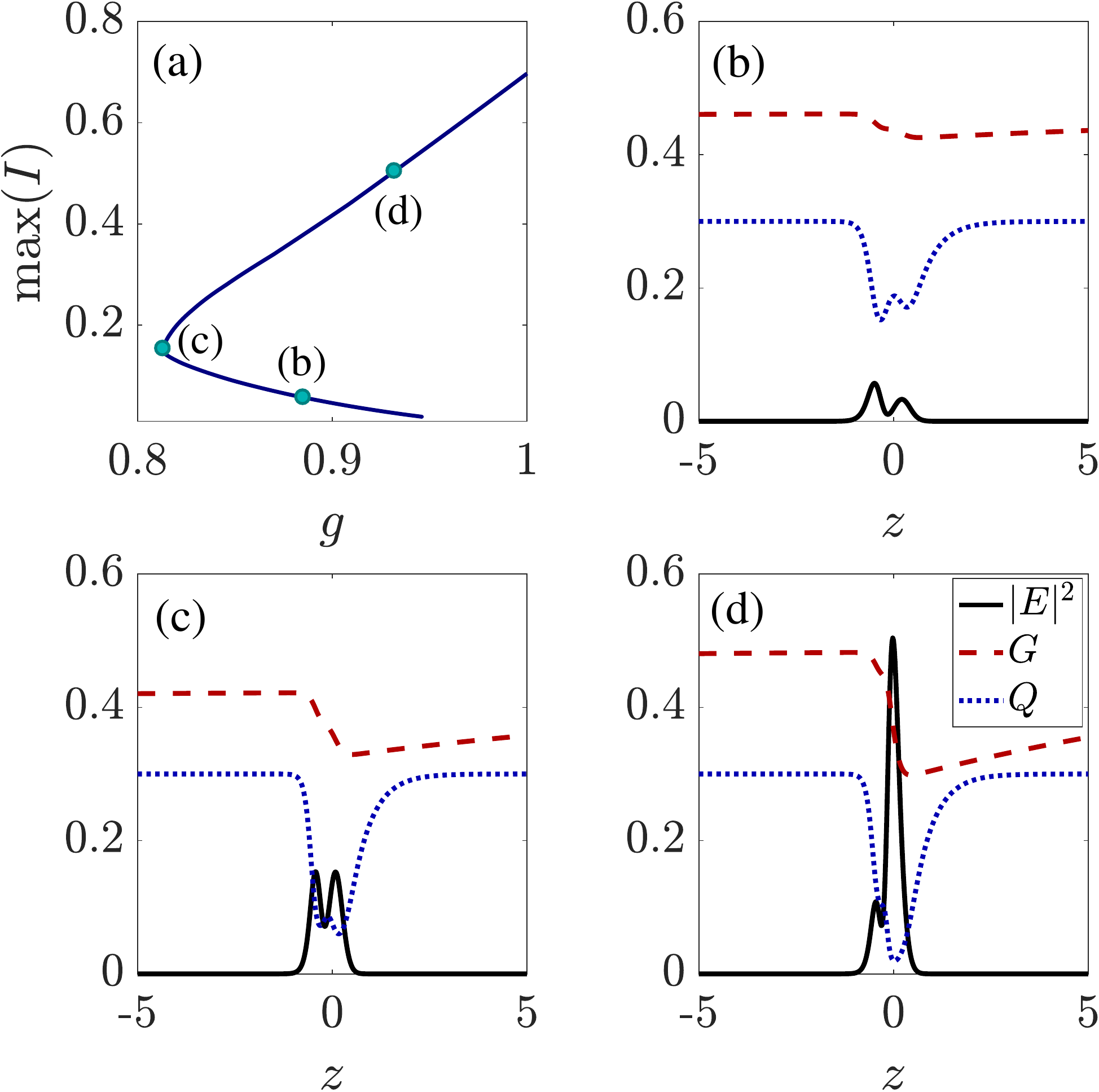}
\par\end{centering}
\centering{}\caption{(Color online) (a) Branch for the two-peaked LS solution obtained
for $(\alpha,\,\beta)=(1.5,\,0.5)$ as a function of the normalized
gain $g$ where we represent the maximum intensity. Temporal profiles
for the low intensity solution (b) at $g_{b}=0.885$, the solution
at the fold $g_{c}=0.813$ (c) and on the upper branch (d) at $g_{d}=0.932$.
The whole solution branch is unstable. Other parameters are the same
as in Fig.~\ref{fig:Haus1}. \label{fig:haus2}}
\end{figure}

In our case, the primary continuation parameter is, e.g., the gain
parameter $g$ or the linewidth enhancement factor $\alpha$. However,
the spectral parameter $\omega$ and the drift velocity $\upsilon$
become two additional free parameters that are automatically adapted
during the continuation. In order to determine $\left(\omega,\upsilon\right)$,
we impose additional auxiliary conditions. In particular, we set the
solution speed, defining $u\left(z,\sigma\right)=\left[\Re\left(E\right),\,\Im\left(E\right),\,G,\,Q\right]$,
by using the integral phase condition 
\begin{eqnarray}
\int u\cdot\frac{\partial u_{\mathrm{old}}}{\partial z}dz & = & 0,
\end{eqnarray}
where $u_{\mathrm{old}}$ denotes the solution obtained in the previous
continuation step. Further, one needs an additional auxiliary condition
to break the phase shift symmetry of the system in order to prevent
the continuation algorithm to trivially follow solutions along the
corresponding neutral degree of freedom. This condition can be easily
implemented by, e.g., setting the phase of the LS to zero in the center
of the computational domain. This condition allows finding the value
of $\omega$ and reads 
\begin{eqnarray}
\Im\left[E\left(\frac{L}{2}\right)\right] & = & 0.
\end{eqnarray}

To increase computational efficiency, we used a domain whose length
$L$ is much smaller than the recovery time of the gain and set $L=10$.
In addition, we impose no-flux boundary conditions on both ends of
the numerical domain 
\begin{equation}
\frac{du}{dz}\left(0\right)=\frac{du}{dz}\left(L\right)=0,
\end{equation}
while the number of mesh points is $N=512$. We note that other kinds
of boundary conditions as, e.g., setting $E\left(0\right)=E\left(L\right)=0$
gave very similar results. Notice that in the case where the domain
is sufficiently large so that if the field intensity is zero, the
proper conditions for $G$ and $Q$ are of the Robin type and are
simply Eqs.~(\ref{eq:H2b}-\ref{eq:H3b}) setting $E=0$
\begin{eqnarray}
\frac{\partial G}{\partial z}+\Gamma G & = & \Gamma G_{0}\,\quad\quad\frac{\partial Q}{\partial z}+Q=Q_{0}.
\end{eqnarray}

Now one can start at, e.g., a numerically given solution, continue
it in parameter space, and obtain a LS solution branch. The result
is depicted in Fig.~\ref{fig:Haus1}, where the evolution of (a)
the (peak) intensity $I$ and (b) the drifting speed $\upsilon$ as
a function of the normalized gain $g$ is presented. We observe that
the main branch of the temporal LS bifurcates from $g=g_{th}=1$,
possesses a fold at some fixed value $g_{SN}$ (marked as the red
circle in Fig.~\ref{fig:Haus1}) and goes to higher intensities.
Note that in the case of Eqs.~(\ref{eq:H1b}-\ref{eq:H3b}), the
solution appears upon increasing $g$ as a saddle-node bifurcation
(SN) and not a saddle-node of limit cycle (SNL) as for Eqs.~(\ref{eq:VT1}-\ref{eq:VT3}).
The critical value is $g_{SN}=0.721$ which compares very well with
the results of the DDE model for which we have $g_{SNL}=0.716$. We
note that the drifting speed $\upsilon$ is a decreasing function
of $g$ for the stable branch of the solution. This result is in good
agreement with the solutions of Eqs.~(\ref{eq:VT1}-\ref{eq:VT3})
because the drift velocity can be identified with the deviation of
the period with respect to $T_{0}$, per unit of $\tau,$ hence the
corresponding transformation is $\upsilon=\left(T-T_{0}\right)$.
Further, in Fig.~\ref{fig:Haus1} (c,d) we show two exemplary stationary
LS profiles that exist for different values of $g$. One can see that
the peak intensity of the LS changes significantly along the branch,
leading to the formation of a narrow peak of high intensity at the
upper branch part.

The Haus PDE~(\ref{eq:H1b}-\ref{eq:H3b}) also predicts the existence
of additional branches of solutions that are composed of several peaks.
We depict in Fig.~\ref{fig:haus2} the secondary branch of two-peaked
solutions. Here, a double peak LS emerges at low intensities and folds
back at $g_{SN}^{\left(2\right)}=0.813$ which compares very well
with $g_{SNL}^{\left(2\right)}=0.808$ given by the DDE model. In
addition, in Fig.~\ref{fig:haus2} (b-d) we depict three exemplary
LS profiles that exist for different values of $g$. As in the case
of the DDE model (cf. Fig.~\ref{fig:2}), the low intensity branch
is composed of two-bumps solutions of different heights and evolve
toward a single bump solution for high values of $g$ at the upper
branch part. For the third branch, we were not able to find a proper
starting solution, as the whole branch is unstable, which, however,
does not mean the three-peaked solution does not exist in Eqs.~(\ref{eq:H1b}-\ref{eq:H3b}).

\begin{figure}
\includegraphics[clip,width=1\columnwidth]{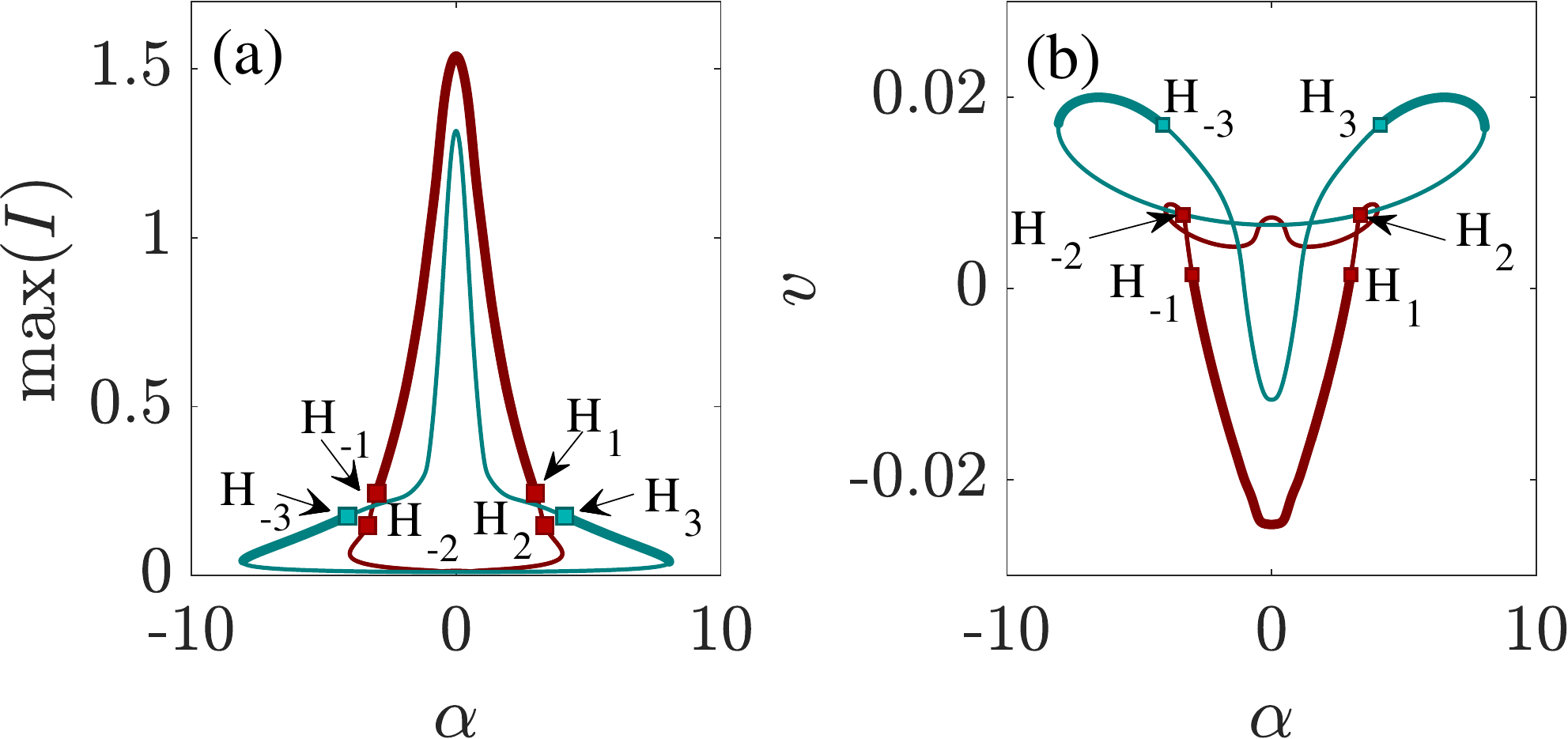}

\caption{(Color online) A bifurcation diagram as a function of $\alpha$ obtained
for $\beta=0$ at a fixed value of the gain $g=0.955$. We represent
(a) the maximum intensity as well as (b) the drifting speed of the
solution $\upsilon$. Stability is indicated with thick lines: On
the main branch (red) the LS is stable between the AH points $\mathrm{H_{\pm1}}$,
the secondary AH bifurcations are indicated as $\mathrm{H_{\pm2}}$.
The LS on the secondary branch (cyan) is stable between the AH points
$\mathrm{H_{\pm3}}$ and the folds. Other parameters are the same
as in Fig.~\ref{fig:Haus1}. \label{fig:SimpleLoopH}}
\end{figure}

In addition to stationary LS solutions, Eqs.~(\ref{eq:H1b}-\ref{eq:H3b})
also predict the existence of temporally oscillating solutions. We
start their analysis with the case where the line enhancement factor
of the absorber $\beta=0$ and perform a continuation in $\alpha$.
There, the branches with different numbers of peaks emerge and reconnect
via the same scenario as in the DDE model~(\ref{eq:VT1}-\ref{eq:VT3})
involving transcritical bifurcations (cf. Fig.~\ref{fig:transLoops})
although it is much more difficult to obtain such results within the
PDE continuation. An example of the resulting branch structure for
$g=0.955$ is depicted in Fig.~\ref{fig:SimpleLoopH}, where the
branches of the primary (red) and secondary (cyan) solutions are shown
after the re-connection. 

\begin{figure}
\centering{}\includegraphics[bb=0bp 0bp 381bp 368bp,clip,width=0.5\columnwidth]{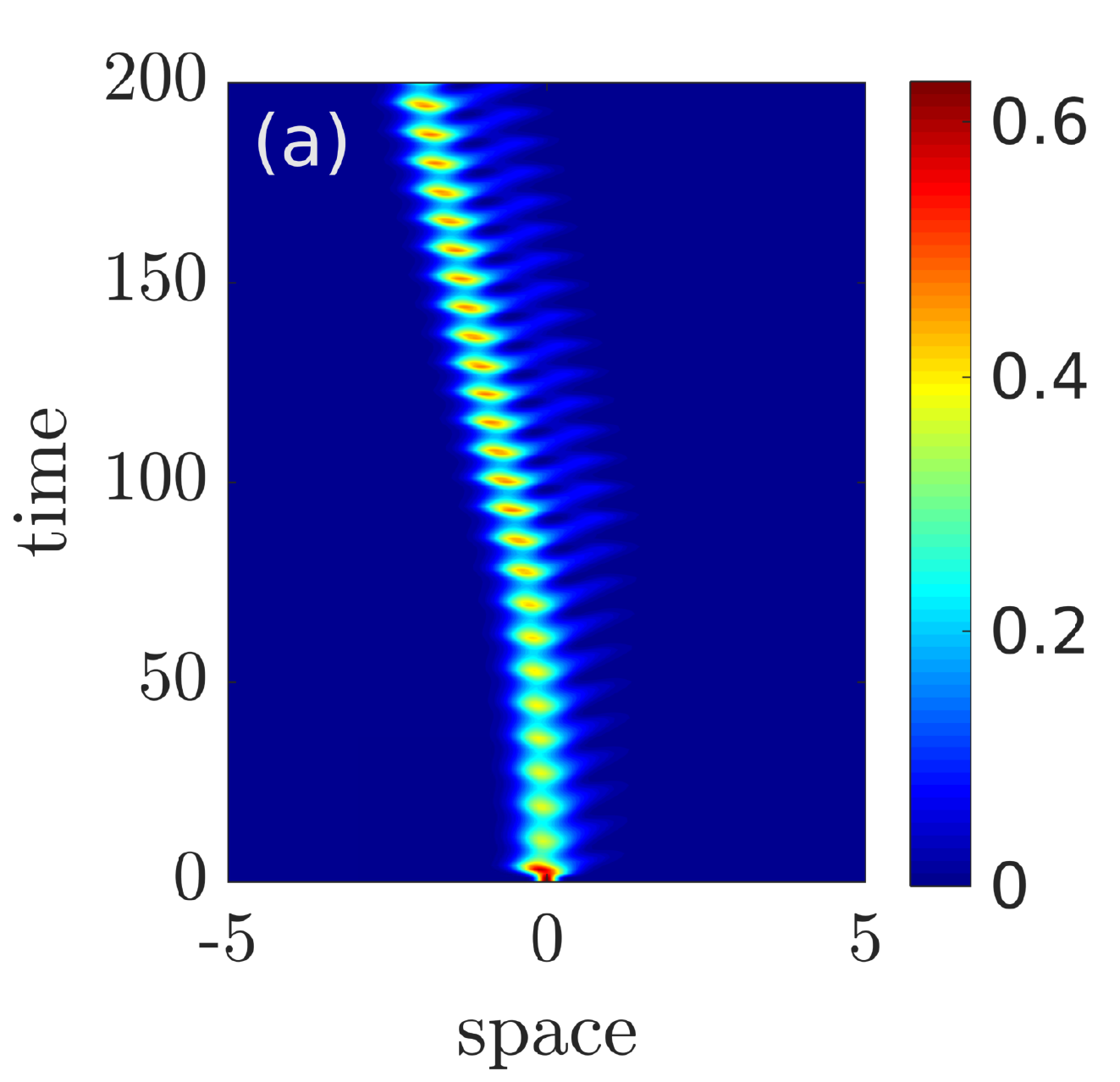}\includegraphics[bb=0bp 0bp 381bp 368bp,clip,width=0.5\columnwidth]{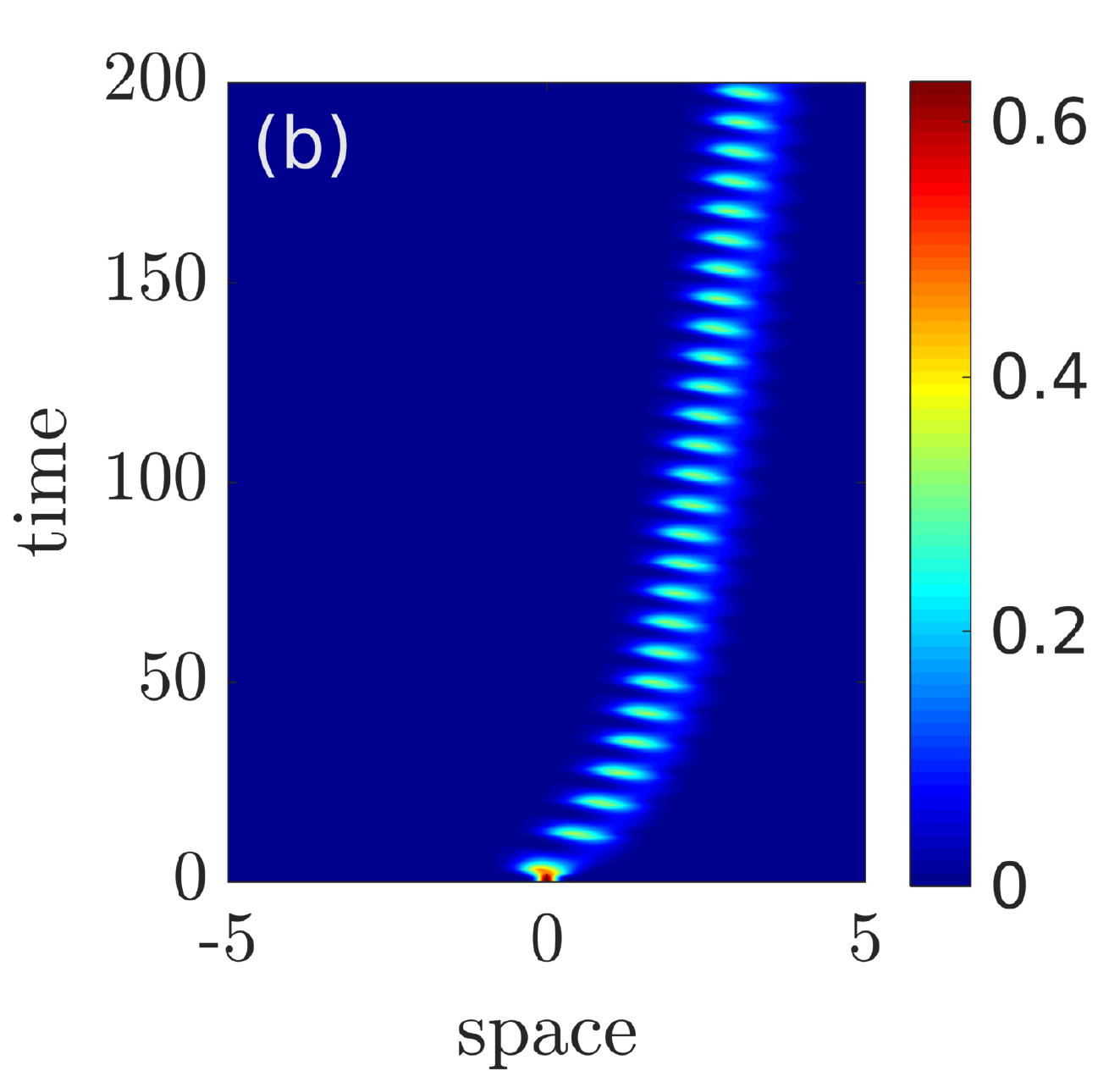}\caption{(Color online) Space-time representation of the intensity field of
AH unstable solutions obtained from direct numerical simulations of
the model~(\ref{eq:H1b}-\ref{eq:H3b}) for different values of $\alpha$
at fixed $g=0.955$ and $\beta=0$ (cf. Fig.~\ref{fig:SimpleLoopH}).
Parameters are chosen to be close to the AH points $\mathrm{H_{1}}$(a)
$\alpha=2.8$ and $\mathrm{H_{3}}$ (b) $\alpha=3.6$. Other parameters
are the same as in Fig.~\ref{fig:Haus1}. \label{fig:AH_ST}}
\end{figure}

One can see that on the main branch the LS is stable between the symmetrically
situated AH bifurcation points $\mathrm{H_{\pm1}}$, whereas the secondary
AH bifurcations appear at $\mathrm{H_{\pm2}}$. Further, the LS solution
on the secondary branch becomes stable for the high $\alpha$ values
between the AH points $\mathrm{H_{\pm3}}$ and the corresponding folds,
which is again in agreement with the DDE results (cf. Fig.~\ref{fig:transLoops}).
In addition, in Fig.~\ref{fig:AH_ST} we show a space-time representation
of the intensity field evolution obtained by direct numerical simulations
of Eqs.~(\ref{eq:H1b}-\ref{eq:H3b}) for two different values of
$\alpha$ close to the AH bifurcation points $\mathrm{H_{1}}$, $\mathrm{H_{3}}$
keeping the other parameters fixed. For the numerical integration
of the model in question a Fourier based semi-implicit split-step
method is employed, see the appendix of \cite{GJ-PRA-17}. Our results
reveal that indeed two AH bifurcations can be found for different
values of $\alpha$ that co-exist at a fixed value of $g$. 

\begin{figure}
\includegraphics[bb=0bp 0bp 577bp 263bp,clip,width=1\columnwidth]{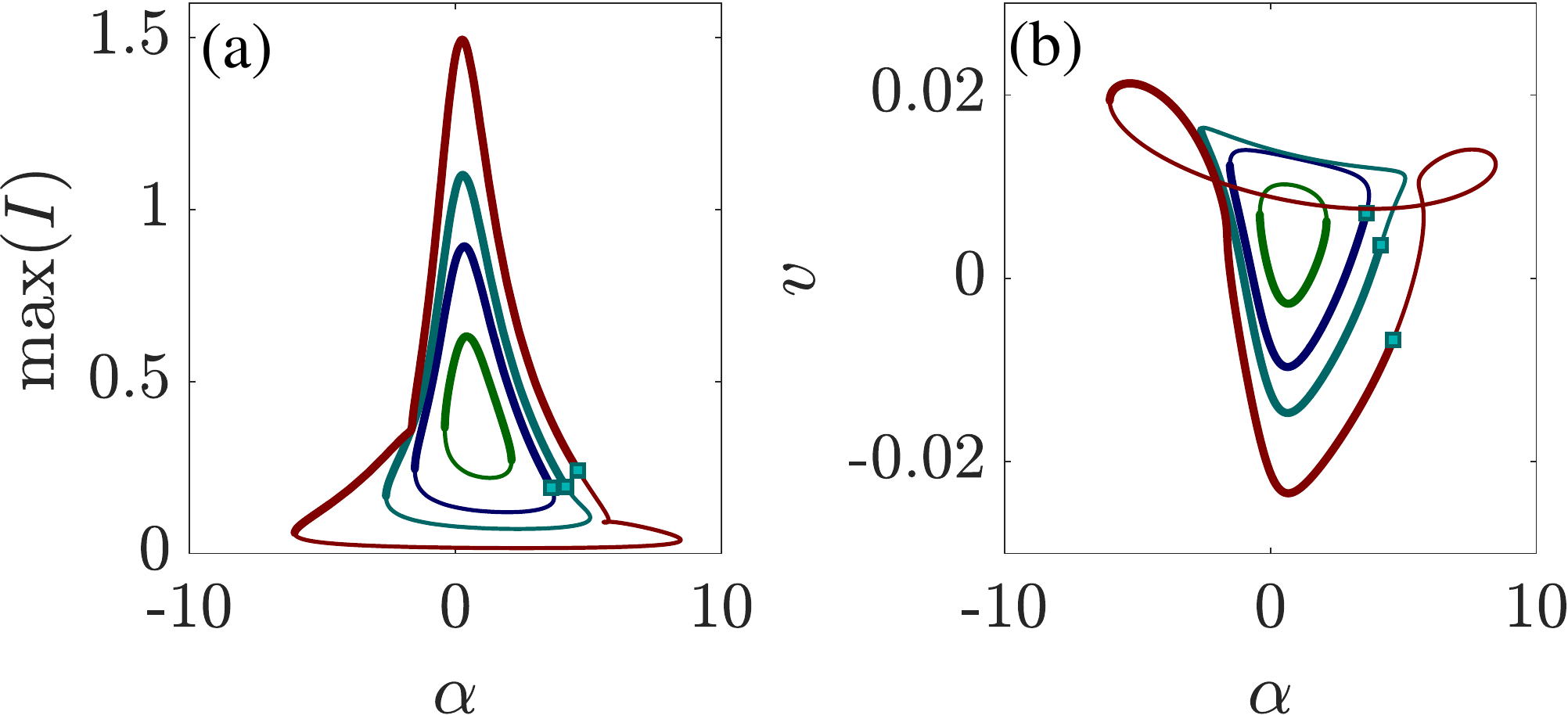}

\caption{(Color online) Two-dimensional bifurcation diagram in $\alpha$ for
$\beta=0.5$ when increasing the gain $g$. We represent the maximum
intensity of the main LS solution (a) as well as its the drifting
speed $\upsilon$ (b). The values of the gain are $g=0.73$ (green),
$g=0.78$ (blue), $g=0.83$ (cyan) and $g=0.94$ (red). Stability
is indicated with thick lines, whereas cyan squares indicate the positions
of the AH bifurcation. Other parameters are the same as in Fig.~\ref{fig:Haus1}.
\label{fig:ComplexLoopH}}

\end{figure}

\begin{figure}
\begin{centering}
\includegraphics[clip,width=1\columnwidth]{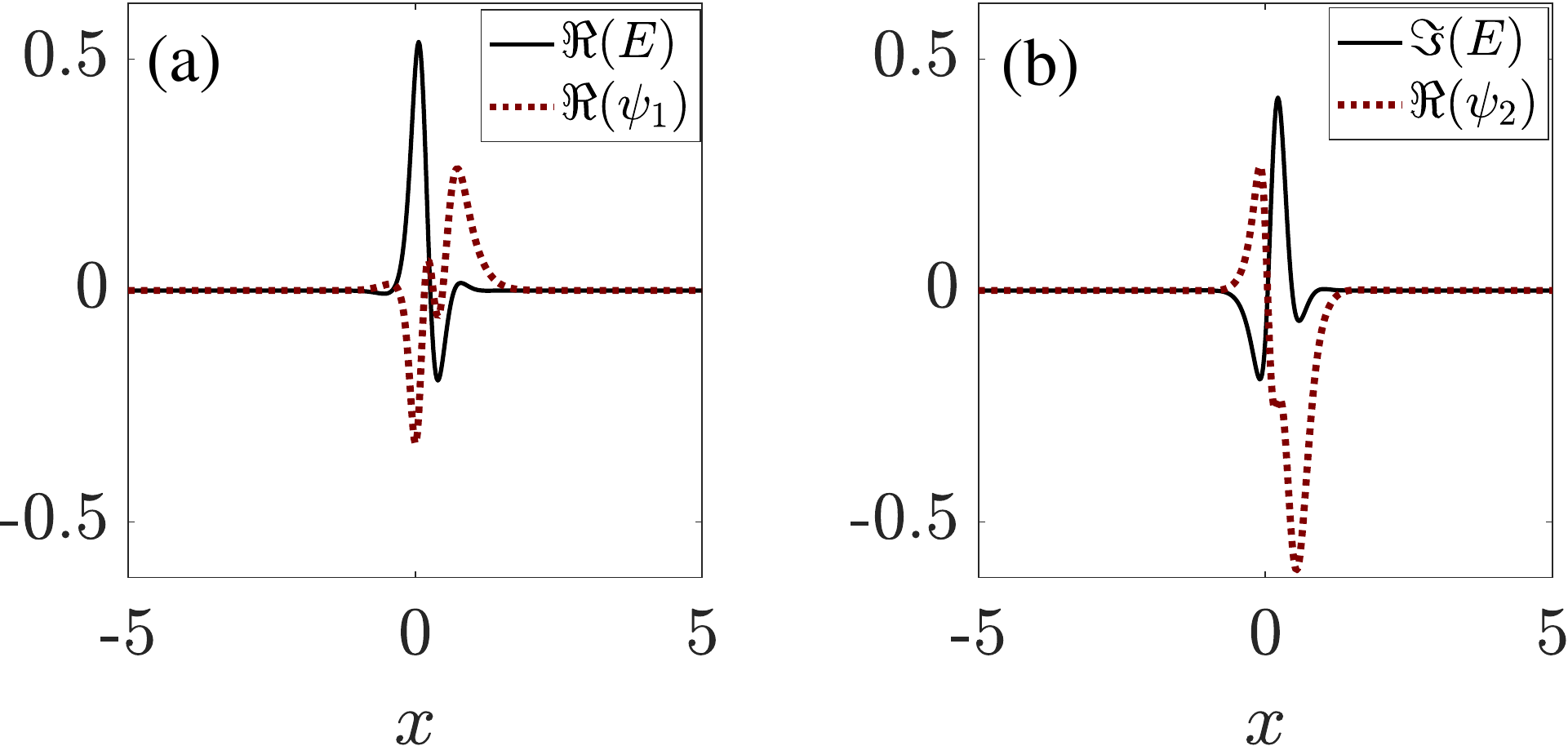}
\par\end{centering}
\centering{}\caption{(Color online) Real parts of the first two components of the critical
eigenfunction $\psi$ of the Hopf unstable solution (red dashed lines)
together with the $(\Re(E),\Im(E))$ components of the field (black
solid lines). Parameters are $\left(g,\alpha,\beta\right)=\left(0.94,4.179,0.5\right)$
(cf. Fig.~\ref{fig:ComplexLoopH}). Other parameters are the same
as in Fig.~\ref{fig:Haus1}. \label{fig:AH_ST-ev}}
\end{figure}

As in the case of the DDE model, for $\beta=0$ the resulting branches
are perfectly symmetrical. However, when $\beta\neq0$ the symmetry
of the diagram is broken and one does see how the solution curves
deform when the gain is increased in Fig.~\ref{fig:ComplexLoopH}.
Here, the evolution of the peak intensity (a) and the drifting velocity
(b) of the main solution branch are presented for $\beta=0.5$. Stability
of the LS solution is indicated with thick lines, whereas cyan squares
mark the positions of appearing AH bifurcations. At variance with
ddebiftool, for the PDE model~(\ref{eq:H1b}-\ref{eq:H3b}) we have
an access to the critical eigenfunctions of the system that inform
on the particular shape of the waveform. An example of the real parts
of the first two components of the critical eigenfunction $\psi=[\psi_{1},\psi_{2},\psi_{3},\psi_{4}]$
associated with the AH instability (dashed red lines) are shown together
with the corresponding $(\Re(E),\,\Im(E))$ components of the field
(solid black lines) in Fig.~\ref{fig:AH_ST-ev}. Here, the parameters
are chosen to be close to the AH bifurcation point at the red line
of Fig.~\ref{fig:ComplexLoopH}, corresponding to $g=0.94$. It turns
out that the components of the unstable eigenfunction are localized
on the trailing edge of the field components. That is, the branch
of the LS gets destabilized via oscillations, localized on the trailing
edge of the LS (cf. Fig.~\ref{fig:AH_ST}). 

\begin{figure}
\begin{centering}
\includegraphics[clip,width=1\columnwidth]{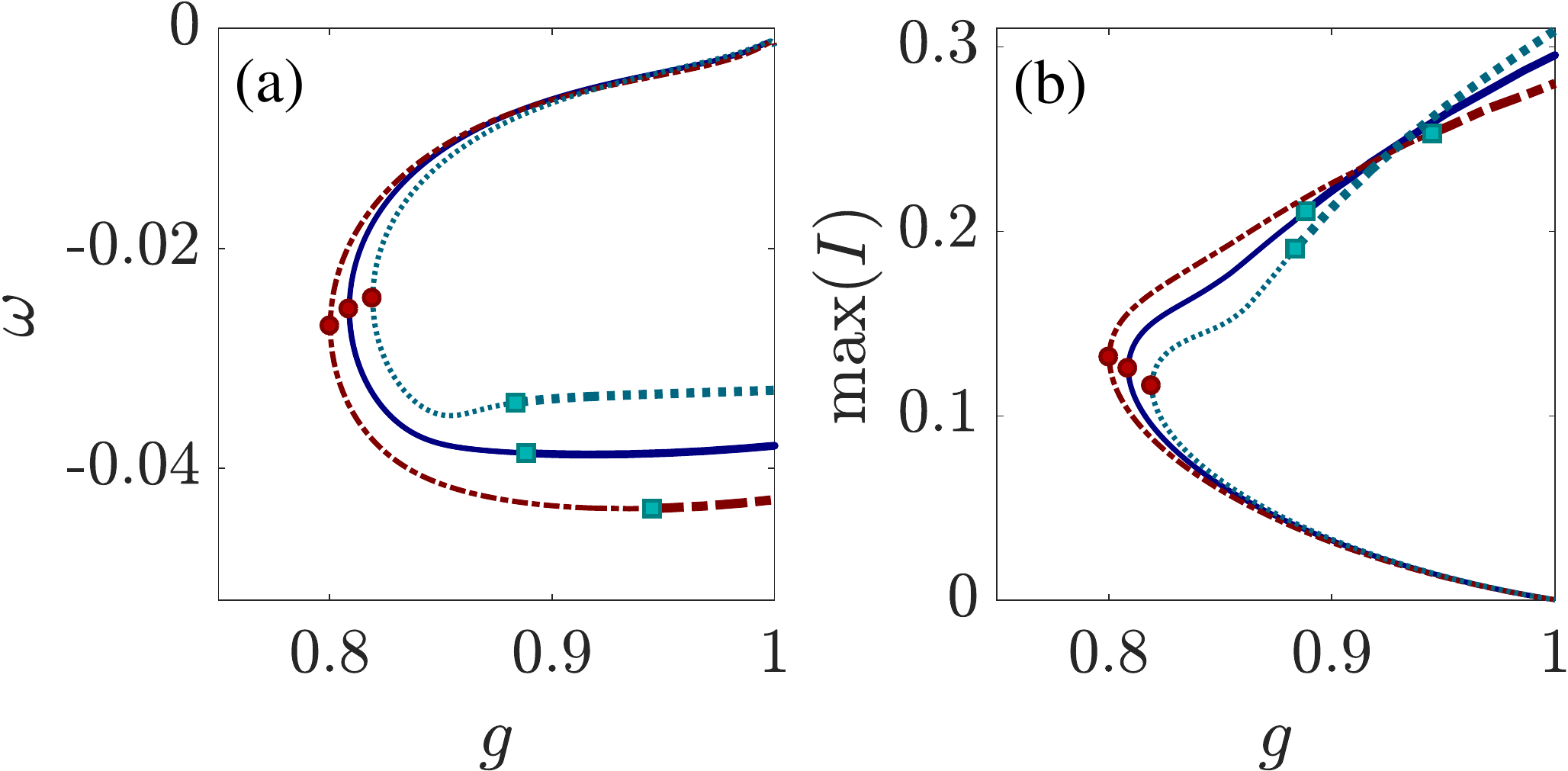}
\par\end{centering}
\begin{centering}
\includegraphics[clip,width=1\columnwidth]{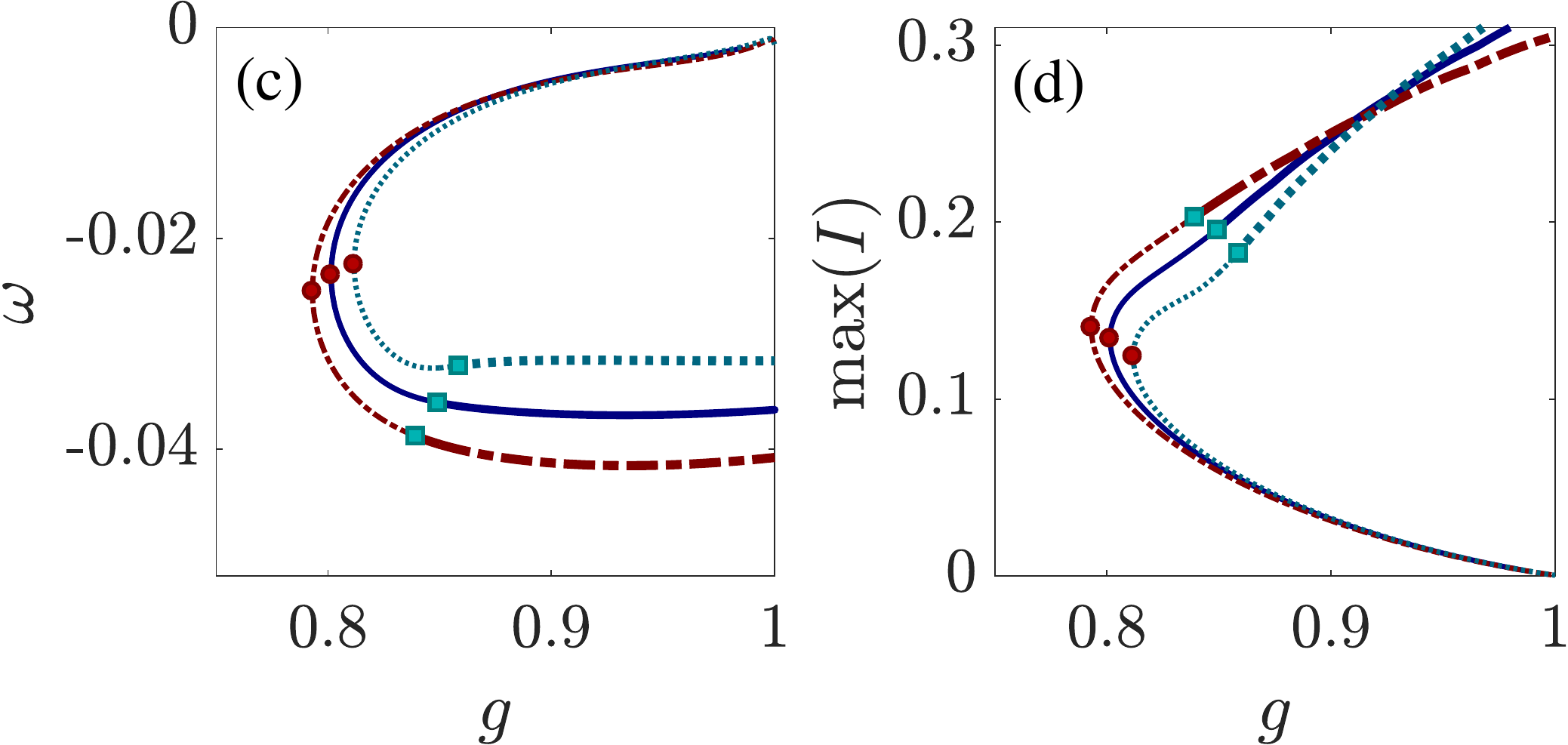}
\par\end{centering}
\centering{}\caption{(Color online) Main solution branch as a function of the normalized
gain $g$ for different amounts of the GVD parameter $D$ and different
values of $\alpha$ at fixed $\beta=0.5$. (a,b) $\alpha=4.5$ and
(c,d) $\alpha=4.3$. We represent the maximum intensity (a,c) and
the frequency shift $\omega$ (b,d). The different curves correspond
to $D=-10^{-3}$ (dash-dotted red), $D=0$ (solid blue) and $D=10^{-3}$(dotted
cyan). The corresponding saddle-nodes (red circles) in (a,b) are located
at $g_{SN}=[0.80,\,0.808,\,0.819]$, while the AH bifurcation positions
(cyan squares) are $g_{AH}=[0.944,\,0.888,\,0.883]$. In (c,d) the
saddle-nodes (red circles) are located at $g_{SN}=[0.793,\,0.801,\,0.811]$,
while the AH bifurcation positions (cyan squares) are $g_{AH}=[0.839,\,0.849,\,0.858]$.
Other parameters are the same as in Fig.~\ref{fig:Haus1}. \label{fig:Hopf_gvd}}
\end{figure}

Interestingly, we can also show how the AH bifurcation can be inhibited
or activated by considering the influence of group velocity dispersion
(GVD). We note that, while this analysis is direct within the framework
of the modified Haus equation, and simply consists in adding an imaginary
contribution to the second order derivative in $z$ in Eq.~(\ref{eq:H1b})
$+iD\partial_{z}^{2}E$, it is not directly possible to do the same
transformation with the DDE model. Adding some amount of dispersion
in a DDE model can only be done via a much more involved method \cite{PSH-PRL-17}.
We note that the dispersion coefficient $D$ corresponds to $D=-\beta_{2}/\tau$
with $\beta_{2}$ the chromatic dispersion. As such $D>0$ corresponds
to anomalous dispersion which favors, e.g., with a self-focusing nonlinearity
$\sim+i\left|E\right|^{2}E$, the appearance of bright solitons. In
our case, however, the effect of GVD is more complex than for the
case of weakly dissipative solitons because the nonlinearity can be
either focusing or defocusing depending on the values of $\alpha$
and $\beta$. In addition the nonlinearity is mediated by two dynamical
variables having very different time scales. To illustrate the influence
of GVD on the LS behavior we show in Fig.~\ref{fig:Hopf_gvd} the
evolution of the main solution branch in $g$ for two different values
of $\alpha$ and three different values of $D$. Here, we represent
the maximum intensity (a,c) and the spectral parameter $\omega$ (b,d).
We notice in Fig.~\ref{fig:Hopf_gvd}(a-b) for $\alpha=4.5$ that
the solution is stable beyond the AH bifurcation (cf. thick lines).
This AH point actually corresponds to the first, subcritical, secondary
AH bifurcation depicted in Fig.~\ref{fig:AH_diag} below which the
oscillation rapidly explodes nonlinearly. Here the effect of positive
GVD is to inhibit the AH bifurcation. Some amount of anomalous dispersion
favors the existence of the temporal LSs as it %
{} pushes the secondary AH bifurcation to higher values of $g$, resulting
in an extended range of stability in Fig.~\ref{fig:Hopf_gvd}(a,b).
Yet, this scenario is changed for slightly smaller value of $\alpha=4.3$,
where one can see that the effect of GVD is inverse and favors the
secondary AH for $D>0$ while inhibits it for $D<0$. From this analysis
we can draw the conclusion that while the main branch characteristics
such as the folding point, intensity and pulse shape are well reproduced
by the exponential Haus master equation, the scenario for the secondary
AH bifurcation is affected. In particular, while we do see the emergence
of the subcritical AH bifurcation, the supercritical AH is absent.
This difference can be ascribed to the fact that the carrier frequency
of the solution $\omega$ oscillates in time leading to a delayed
phase $\omega\tau$ that is slowly evolving, a feature lost in the
PDE mapping presented in Eqs.~(\ref{eq:H1b}-\ref{eq:H3b}).%

\begin{figure}
\begin{centering}
\includegraphics[clip,width=0.8\columnwidth]{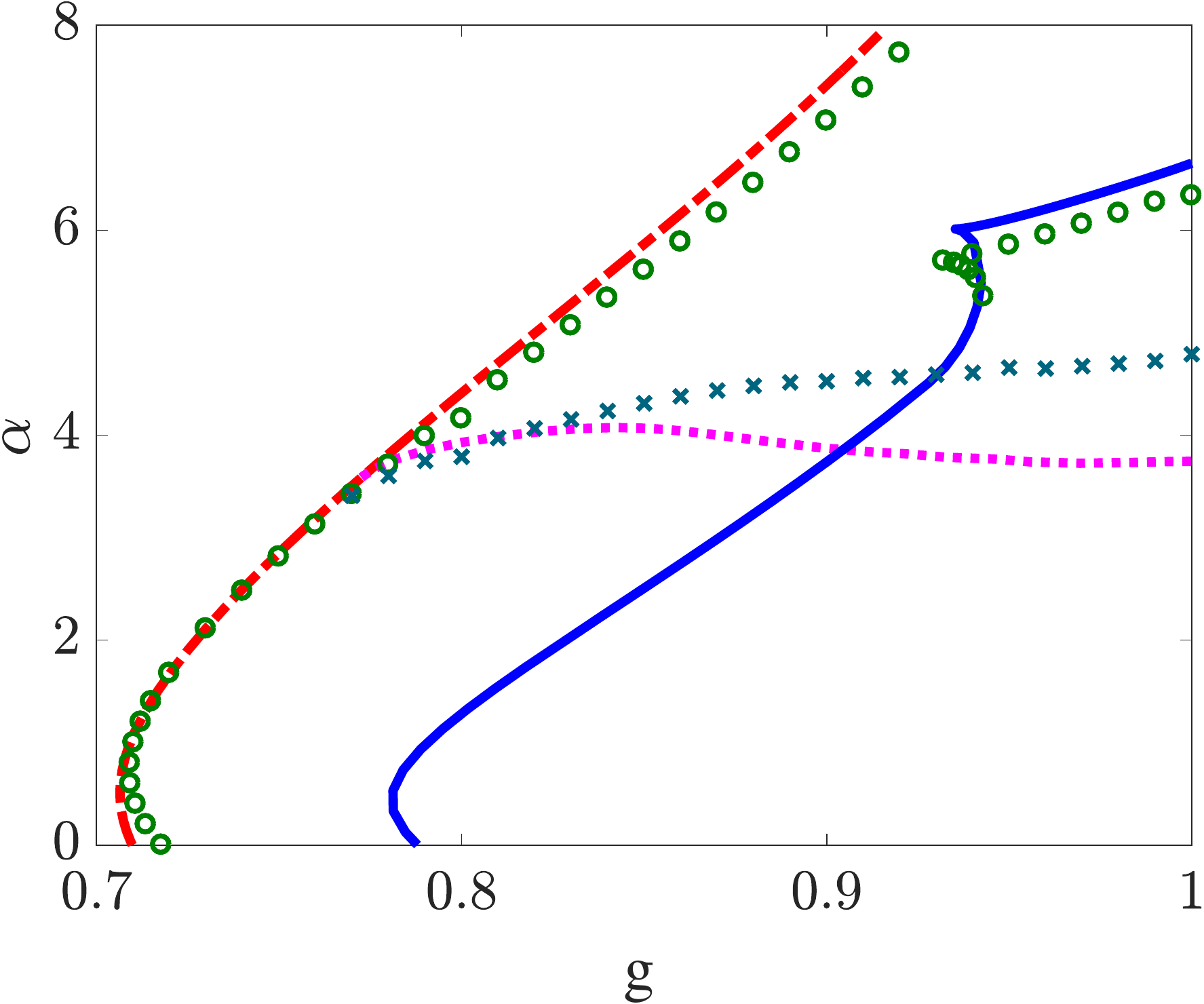}
\par\end{centering}
\centering{}\caption{(Color online) Bifurcation diagram in the $\left(g,\alpha\right)$
plane, showing the evolution of the bifurcation points of both DDE
and PDE models . The primary SNL bifurcation (dotted red line) of
the DDE defines the breadth of the paraboloid for the primary solution.
The secondary SNL is depicted in solid blue and the cusp is visible
around $g=0.935$. The secondary AH bifurcation is depicted in dotted
pink and it connects with the SNL on a codimension two point. The
SN points of the PDE are shown as green circles, whereas cyan crosses
stay for the primary AH points. The cusp for the PDE is around $g=0.932.$
Other parameters are $\beta=0.5$. \label{fig:bifdiag_dde}}
\end{figure}

Finally, we depict the summary of our bifurcation analysis of both
the DDE and the PDE models in Fig.~\ref{fig:bifdiag_dde} allowing
for a more direct global comparison. Here we represent the bifurcation
diagram in the $(g,\,\alpha)$ plane, showing the SNL points of the
DDE for both the primary (red dashed line) and the secondary (solid
blue line) branches, and compared it with the SN points of the PDE
(green circles), as well as the the secondary AH bifurcation occurring
on the primary branch in the DDE (pink dotted line) in addition to
the primary AH of the PDE model (cyan crosses). Here, the appearance
of the cups is visible for both models. We do notice a small deviation
for the folding point of the solution while the secondary Andronov-Hopf
lines are significantly different. While the AH lines grows and falls
as a function of $g$ in the DDE case, the one in the PDE model is
steadily increasing, which explains the difference encountered in
Fig.~\ref{fig:Hopf_gvd}. While scanning $g$ in the DDE model, one
can cross twice the AH line, giving rise to the sub- and supercritical
limit cycles depicted, e.g., in Fig.~\ref{fig:3}(c,d), the line
can only be crossed once in the PDE model, giving rise only to the
subcritical limit cycle. Finally, it was not possible to follow the
cusp bifurcation on the secondary branch in the PDE model for all
values of $g$, although we believe that it would closely follow the
same trend as in the DDE model.

\section{Conclusion}

In conclusion, we discussed the bifurcation and the stability analysis
of the time periodic solutions of the PML laser found in the long
delay limit. We demonstrated that besides the main solution branch
disclosed in \cite{MJB-PRL-14}, numerous additional branches exist,
and that upon increasing the bias current, they splice with the main
solution loop via transcritical bifurcations leading to a seemingly
self-intersecting manifold for the solutions. We showed that for large
but realistic values of the $\alpha$-factor in the gain section,
the secondary branch is an essential part of the bifurcation scenario
as it is the one giving the stable solution for large gain. A secondary
Andronov-Hopf (AH) bifurcation is found either increasing the gain
or the $\alpha$ factor leading to slowly evolving oscillations of
the LS waveform. As the destabilization is found for increasing $\alpha$
factor, this points toward a dispersive nature of this instability.
In addition, the bifurcation analysis of the modified Haus equation
is presented. We showed that this model needs to consider an additional
phase factor in order to properly reproduce the lasing threshold.
A good agreement was found, not only for a single time trace, but
for the whole bifurcation diagram, although the analysis with pde2path
proved to be more technically involved. It was shown that the Andronov-Hopf
instability found for large $\alpha$-values can be mitigated by introducing
some amount of group velocity dispersion which counteracts the dispersive
effect induced by the material. Our preliminary study indicated that
GVD may have a profound impact on the dynamics of temporal LSs. Notice
that introducing GVD at the level of the Haus PDE is direct while
it is known to be quite challenging in the DDE approach.

While we found a good agreement for the bifurcation diagram explaining
the emergence of the single LS, we found some discrepancies regarding
the secondary instabilities. In particular, the evolution of the secondary
AH line in the $\left(g,\alpha\right)$ plane was found to be significantly
different, leading to a quantitatively different bifurcation scenario
for values of $\alpha$ in a particular interval: While the AH line
could be crossed twice in the DDE model, it is only crossed once in
the ``equivalent'' PDE. However, this discrepancy was found to occur
only in a small interval of the linewidth enhancement factor of the
gain section. Overall we demonstrated in this manuscript that, while
the coherent modal structure of the DDE is lost due to the absence
of boundary conditions and the secondary AH regime can be shifted,
the exponential Haus master equation can be considered as an effective
order parameter equation representing the dynamics of a temporal LS
found in the DDE model. The good agreement between the two approaches
validates further studies regarding the effect of GVD on temporal
LSs, but also on light bullets. 
\begin{acknowledgments}
We acknowledge financial support project COMBINA (TEC2015-65212-C3-3-P
AEI/FEDER UE) and the Ramón y Cajal fellowship. S.V.G. acknowledges
the Universitat de les Illes Balears for funding a stay where part
of this work was developed.
\end{acknowledgments}

\section*{Appendix}

Analytical solutions for the pulse shape in the subcritical region
below threshold can only be found in the so-called Uniform Field Limit
(UFL) where the gain, absorption and losses are small at each round-trip.
We note that these approximations mean that $G$ and $Q$ are small,
but their responses are not necessarily weakly nonlinear in the field
intensity. The UFL consists in linearizing the gain and absorption
per pass in Eqs.~(\ref{eq:H1b}-\ref{eq:H3b}), setting $e^{G}=1+G$
and $e^{-Q}=1-Q$. This approximation will allow us to factor out
the cavity losses. To do, so we define the new expression of the threshold
in this linearized model as
\begin{eqnarray}
G_{th} & = & Q_{0}+2\frac{1-\sqrt{\kappa}}{\sqrt{\kappa}}\,.
\end{eqnarray}
We also defined the normalized absorption as
\begin{eqnarray}
q & = & \frac{\sqrt{\kappa}Q}{2\left(1-\sqrt{\kappa}\right)}
\end{eqnarray}
and as such, $g=G/G_{th}$, $g_{0}=G_{0}/G_{th}$ and $q_{0}=\sqrt{\kappa}Q_{0}/\left[2\left(1-\sqrt{\kappa}\right)\right]$
leading to
\begin{eqnarray*}
G_{0} & = & 2g_{0}\frac{1-\sqrt{\kappa}}{\sqrt{\kappa}}\left(q_{0}+1\right)\,.
\end{eqnarray*}
Replacing these expressions into the linearized Eq.~(\ref{eq:H1b}),
we find the following, 
\begin{eqnarray}
\frac{\partial E}{\partial\tilde{\sigma}}-\frac{1}{2\tilde{\gamma}^{2}}\frac{\partial^{2}E}{\partial z^{2}}&=&\left[\left(1-i\alpha\right)\left(q_{0}+1\right)g- \right. \\ \nonumber
    &&\left. -\left(1-i\beta\right)q-1+i\theta\right]E,\label{eq:Hl1}\\
\frac{\partial g}{\partial z} & = & \Gamma\left(g_{0}-g\right)-g\left|E\right|^{2},\label{eq:Hl2}\\
\frac{\partial q}{\partial z} & = & q_{0}-q-qs\left|E\right|^{2},\label{eq:Hl3}
\end{eqnarray}
where we normalized the slow time as $\tilde{\sigma}=\left(1-\sqrt{\kappa}\right)\sigma$,
the filter bandwidth $\tilde{\gamma}=\sqrt{1-\sqrt{\kappa}}$$\gamma$
and the phase $\Theta$ as $\theta=\Theta\sqrt{\kappa}/\left(1-\sqrt{\kappa}\right)$
hence $\theta=\alpha\left(q_{0}+1\right)g_{0}-\beta q_{0}$. In Eqs.~(\ref{eq:Hl1}-\ref{eq:Hl3}),
the parameter $\kappa$ is now factored out, and the non-saturable
losses are unity. Also, the lasing threshold is now conveniently $g=1$.
Dimensional analysis indicates that the pulse-width is typically $\tau_{p}\sim1/\tilde{\gamma}$
and the pulse peak intensity $\sim\tilde{\gamma}$, so that we can
distinguish between the regimes of a slow absorber (found for short
pulses) and that of a fast absorber depending if $\tilde{\gamma}\gg1$
or $\tilde{\gamma}\ll1$. In the first and second cases, the dynamics
of $q\left(z\right)$ are respectively 
\begin{eqnarray}
q_{slow}\left(z\right) & \simeq & q_{0}\exp\left(-s\int_{0}^{z}I\left(u\right)du\right),\\
q_{fast}\left(z\right) & \simeq & \frac{q_{0}}{1+sI\left(z\right)}.
\end{eqnarray}

We search for solutions in the slow absorber regime as the bistable
region below threshold can be found more easily in this regime. We
denote the partially integrated pulse energy $P\left(z\right)=\int^{z}I\left(z,t\right)dz$.
During the pulse emission, the fast stage in which stimulated terms
are dominant, we have
\begin{eqnarray}
g\left(z\right) & = & g_{0}\exp\left[-P\left(z\right)\right]\,,\,q\left(z\right)=q_{0}\exp\left[-sP\left(z\right)\right].\label{eq:solgq}
\end{eqnarray}
We note $P\left(+\infty\right)=P$, the total pulse energy. If, for
the sake of simplicity, we set $\alpha=\beta=0$, the solutions of
Eqs.~(\ref{eq:Hl1}-\ref{eq:Hl3}) are unchirped, drifting hyperbolic
secants of the form 
\begin{eqnarray}
E\left(z,\tilde{\sigma}\right) & =\sqrt{\frac{P}{2\tau}}\mathrm{sech}\left(\frac{z-\upsilon\tilde{\sigma}}{\tau}\right) & .
\end{eqnarray}
Expanding $g\left(z\right)$ and $q\left(z\right)$ in Eq.~(\ref{eq:solgq})
up to second order in $P\left(z\right)$ and identifying the constant,
$\tanh\left(x\right)$ and $\tanh^{2}\left(x\right)$ terms allows
finding a system of equations defining the pulse parameters $\left(P,\tau,\upsilon\right)$
as 
\begin{eqnarray}
0 & = & 2+\left[-4+g_{0}\left(4-2P+P^{2}\right)\left(1+q_{0}\right)\right.\nonumber \\
 & - & \left.q_{0}\left(4-2Ps+P^{2}s^{2}\right)\right]\tilde{\gamma}^{2}\tau^{2}\,,\\
0 & = & 4\upsilon-P\left[g_{0}\left(P-2\right)\left(1+q_{0}\right)+q_{0}s\left(2-sP\right)\right]\tau\\
0 & = & g_{0}P^{2}\left(1+q_{0}\right)+P^{2}q_{0}s^{2}+\frac{8}{\tilde{\gamma}^{2}\tau^{2}}\,.
\end{eqnarray}
Solving the power $P$ as a function of the gain leads to 
\begin{eqnarray}
g_{H}\left(P\right) & = & \frac{16\left(1+q_{0}\right)-8Pq_{0}s+3P^{2}q_{0}s^{2}}{\left(16-8P+3P^{2}\right)(1+q_{0})}\,.\label{eq:g_Haus}
\end{eqnarray}

On the other hand, assuming a Dirac pulse shape $E\left(z,\tilde{\sigma}\right)=\sqrt{P}\delta\left(z\right)$
leads to another solution for the pulse power, in which we neglect
the effect of pulse filtering as given by the second derivative in
Eq.~(\ref{eq:Hl1}) but where we do not need to expand Eq.~(\ref{eq:solgq})
up to second order in $P$. One can see for instance \cite{GJ-PRA-17}
for the details of these calculations, that can also be obtained out
of the UFL as in \cite{VT-PRA-05}. We find the following expression
for the gain as a function of the pulse energy,
\begin{eqnarray}
g_{N}\left(P\right) & = & \frac{\left(1-e^{-Ps}\right)q_{0}+Ps}{\left(1-e^{-P}\right)(1+q_{0})s}.\label{eq:g_New}
\end{eqnarray}

The comparison between the results given by Eq.~(\ref{eq:g_Haus})
and Eq.~(\ref{eq:g_New}) is given in Fig.~\ref{fig:New_vs_Haus}.


\end{document}